\pdfoutput=1
\documentclass[fleqn,usenatbib]{mnras}

\usepackage{newtxtext,newtxmath}
\usepackage{csvsimple}
\usepackage{hyperref}
\usepackage{graphicx}
\usepackage{rotating}
\usepackage{pdflscape}
\usepackage{longtable}
\usepackage{xcolor}
\usepackage{cancel}
\usepackage{orcidlink}
\usepackage{ulem}
\usepackage[T1]{fontenc}
\usepackage{dot2texi}
\usepackage{tikz}
\usetikzlibrary{shapes,arrows}
\DeclareRobustCommand{\VAN}[3]{#2}
\let\VANthebibliography\thebibliography
\def\thebibliography{\DeclareRobustCommand{\VAN}[3]{##3}\VANthebibliography}

\usepackage{graphicx}	
\usepackage{amsmath}	

\title[Infrared highly variable YSOs]{A population of mid-infrared large-amplitude variable young stellar objects from unTimely}

\author[Li \& Wang]{
Jiaxun Li$^{1}$\thanks{E-mail: lijiaxun@mail.ustc.edu.cn}~\orcidlink{0009-0003-5573-9240} and
Tinggui Wang$^{1,2}~\orcidlink{0000-0002-1517-6792}$
\\
$^{1}$Department of Astronomy, University of Science and Technology of China, Hefei, 230026, People’s Republic of China\\
$^{2}$Deep Space Exploration Laboratory, University of Science and Technology of China, Hefei 230026, People’s Republic of China\\
}

\date{Accepted XXX. Received YYY; in original form ZZZ}

\pubyear{2024}

\begin{document}
\label{firstpage}
\pagerange{\pageref{firstpage}--\pageref{lastpage}}
\maketitle

\begin{abstract}
Utilizing a decade-long unTimely dataset, supplemented by multi-band data from archives, we search for young stellar objects (YSOs) with variations larger than one magnitude in W1 band within a region of 110 square degrees in the Galactic plane, covered by VISTA Variables in the Via Lactea (VVV).  
A total of 641 candidate YSOs have been identified. 
We classified them into bursts, dips, faders, seculars, and irregulars.
Within the burst category, 18 sources were identified as FUor candidates and 1 as an EXor candidate. Irregulars are the most prevalent in the sample. 
In both bursts and faders, the redder sources tend to show a pattern of bluer when brighter, whereas the bluer sources display the opposite trend, possibly related to the accretion structure of YSOs at different stages.
Finally, we obtained the recurrence time scale for FUor eruptions at various stages of YSO evolution. 
Our findings indicate that younger YSOs generally experience more frequent eruptions compared to older ones.
\end{abstract}

\begin{keywords}
stars: pre-main-sequence -- infrared: stars -- stars: AGB and post-AGB -- stars: variables: T Tauri, Herbig Ae/Be
\end{keywords}



\section{Introduction} \label{Introduction}

Young Stellar Objects (YSOs) are stars in the pre-main sequence stage of evolution. These objects typically consist of a central star, surrounded by an accretion disk \citep{1973A&A....24..337S,1981ARA&A..19..137P,1987ApJ...323..714K,1997ApJ...490..368C,1998ApJ...500..411D,2001ApJ...560..957D} and possibly an outer envelope \citep{1976ApJ...210..377U,1984ApJ...286..529T,Hartmann_1998}. In addition to the thermal radiation emitted by stars as a result of self-contraction and burning of deuterium, and possibly lithium, YSOs may also emit ultraviolet to X-ray photons due to shock waves generated by the infalling of accreted gas along the magnetic field of the star \citep{1991ApJ...370L..39K,Hartmann_1998,2012A&A...541A.116A,2012ApJ...744...55A}. Active radiation is generated by the accretion disk as a result of viscous dissipation \citep{1973A&A....24..337S,1988ApJ...325..231K,1999MNRAS.308..147G,2007ApJ...669..483Z}, while passive radiation is the reprocessed radiation of the central star by the disk\citep{1997ApJ...490..368C}. The outer envelope reprocesses the radiation from the central star and disk into longer wavelengths. The presence of both a disk and an envelope in YSOs gives rise to excessive infrared radiation over stellar photosphere emission \citep{2006ApJS..167..256R,2007ApJS..169..328R,2017A&A...600A..11R}.

YSOs display a wide range of variations in terms of timescales and amplitudes. These variations can occur over periods as short as a few hours or as long as several decades. In terms of amplitude, YSOs can exhibit fluctuations ranging from a few tenths of a magnitude to 5-7 magnitudes in the optical range \citep{2015ApJ...808...68H,2023ASPC..534..355F}. Variability has been observed across a wide range of wavebands, ranging from X-rays to millimeter-wave bands \citep{1994AJ....108.1906H,2001AJ....121.3160C,2011ApJ...733...50M,2012A&A...548A..85F,Tofflemire_2017,2019ApJ...871...72M,Lee_2021}. Numerous processes may contribute to the variability observed in YSOs, such as surface flares on the star \citep{Vievering_2019}, the existence of both cool and hot spots on the star's surface leading to fluctuations in observed brightness during stellar rotation \citep{1992ApJ...398L..61A,1996AJ....111..283C}, alterations in the rate of accretion \citep{2013MNRAS.431.2673K,2014AJ....147...83S} or obscuration along the line of sight \citep{2013A&A...557A..77B}, and modifications in the properties of the disk \citep{1997AJ....114..288M,1997ApJ...491..885N,2007A&A...463.1017B,2013ApJ...773..145W,2013A&A...557A..77B,2015AJ....150...32R}.

In the early phase of the YSO variability study, the primary focus was on the optical band \citep{1994AJ....108.1906H,2013ApJ...768...93F}. Nevertheless, in recent times, there has been growing interest in the exploration of infrared bands for the study of YSOs owing to advances in large infrared array detectors. The YSOVAR program \citep{2011ApJ...733...50M,2014AJ....148..122G,2014AJ....147...82C,2018AJ....155...99W} has played a crucial role in uncovering widespread infrared variability among YSOs. For instance, \citet{2018AJ....155...99W} discovered that embedded stars exhibit larger variability amplitudes using this program. Furthermore, \citet{2014AJ....148..122G} found that younger stars tend to exhibit longer timescales of variability. \citet{2017MNRAS.465.3011C} conducted a survey of YSOs in a region of 119 square degrees of the Galactic plane using the dataset derived from VISTA Variables in the Via Lactea (VVV, \citealp{2010NewA...15..433M}). By analyzing the K$_\mathrm{s}$ band light curves, they classified these YSOs into eruptives, faders, dippers, short-term variables, long-term periodic variables, and ecliping-binaries. In another study, \citet{2021ApJ...920..132P} identified 1734 variable sources within known YSOs and classified them into linear, curved, periodic, burst, drop, and irregular groups. 
\citet{2019ApJ...883....6U} shifted their attention to massive protostars and successfully identified five variable YSOs from 809 massive protostars in NEOWISE survey.
Recently, using VVV data, \citet{2024MNRAS.528.1789L} discovered 222 high-amplitude variables in Galactic bulge and disc, 40 of which were considered as YSOs.

Among variable YSOs, a distinctive category, called outbursts, exhibited significantly large variability amplitudes \citep{1996ARA&A..34..207H,2014prpl.conf..387A,2023ASPC..534..355F}. These outbursts are believed to arise from episodic accretion. This class encompasses two subtypes: EXors and FUors. The former lasts for months to years, while the latter may persist for several years to a century. These two classes also manifest distinct spectroscopic characteristics. EXors exhibit recombination lines of \ion{H}{i} and \ion{He}{i}, along with metallic collisional excitation lines, as well as CO emission \citep{2017ApJ...839..112G}. On the contrary, FUors show pronounced absorption lines, including CO, \ion{Na}{i}, and $\rm{H}_2\rm{O}$ \citep{2018ApJ...861..145C}. In recent years, with the expansion of observational samples, a new category of outbursts distinct from FUors and EXors has been identified. Their spectral types display a blend of characteristics from FUors and EXors. These sources are classified as MNors \citep{2017MNRAS.465.3011C,2017MNRAS.465.3039C}.

Episodic accretion is considered as a potential solution to the issue of insufficient accretion rates in YSOs, commonly referred to as the luminosity problem \citep{1995ApJS..101..117K,2010ApJ...710..470D,2011ApJ...736...53O,2012ApJ...747...52D,2014prpl.conf..387A}. In this context, it becomes crucial to statistically assess the probability that outbursts occur in YSOs or, in other words, to calculate the recurrence timescales of these events. \citep{2019ApJ...872..183F,2021ApJ...920..132P}. In the past century, dozens of eruptive YSOs have been discovered. However, most of them have been identified in the optical band, which is biased against deeply embedded objects. Therefore, the importance of infrared surveys for YSOs cannot be overstated \citep{2020MNRAS.499.1805L,2023ApJ...957....8W}. These surveys play a crucial role in identifying outburst sources that may go unnoticed in the optical band, while also enhancing our understanding of the source of these eruptions. \citet{2013MNRAS.430.2910S} conducted a study to identify YSOs that exhibit infrared outbursts. They compared the magnitude differences in the infrared data between two epochs obtained from Spitzer and WISE. Their analysis resulted in the discovery of three FUor candidates and one EXor candidate in their sample. \citet{2019ApJ...872..183F} used a similar approach, but also included the 24 $\mu$m band, to identify two outbursting YSOs in the Orion region. \citet{2021ApJ...920..132P} collected W2 light curves for 5398 confirmed YSOs and found 2-9 FUors. \citet{2022ApJ...924L..23Z} utilized Spitzer photometry data with a time baseline of 13 years and identified five outburst events in Orion, two of which were previously unknown. All of these studies offer useful constraints on the recurrence timescale of outbursts in YSOs. However, because of the limited sampling or depth of observations on the light curve and only a few identified outbursts, additional investigation is needed to improve the accuracy of these constraints on the timescales between outbursts. Therefore, further research on the infrared properties of YSOs is crucial.

In this study, we conducted a thorough investigation of variable sources in the sky region $295^\circ<l<350^\circ$ and $-1^\circ<b<1^\circ$, utilizing data from the unTimely W1 catalog. This catalog covers a period of approximately 10 years and provides deeper photometry compared to the NEOWISE catalog \citep{2011ApJ...743..156M,2014ApJ...792...30M}. We identified a total of 641 variable YSOs in the W1 band, with variations exceeding 1 magnitude. We discuss these findings in detail. Section \ref{Mid-infrared variables selection} introduces the archival data used in this study and explains the methodology employed to identify variable sources. In Section \ref{YSO selection}, we describe our approach to identifying YSOs and classifying them into different evolutionary stages based on spectral indices. Section \ref{Lightcurve classify} outlines our classification of these sources based on the morphology of their light curves. In Section \ref{Result}, we analyze variations in the morphology of light curves and amplitudes of YSOs at different evolutionary stages. We also investigate color changes among different types of variable source and constrain the recurrence timescales of FUor objects. Section \ref{Discussion} provides a comparative analysis of our research findings with those of other studies. We also discuss the mechanisms underlying the light variations. Finally, in Section \ref{Conclusions}, we provide a comprehensive summary of the key points presented in this paper.

\section{Mid-infrared variables selection} \label{Mid-infrared variables selection}
This paper focuses on the sky area: $295^\circ<l<350^\circ$ and $-1^\circ<b<1^\circ$, which is also covered by the VISTA Variables in the Via Lactea (VVV) project.

\subsection{WISE}
The Wide Field Infrared Survey Explorer (WISE,\citealp{2010AJ....140.1868W}) is a NASA infrared astronomy space telescope in the Explorers Program. It carried out a whole sky survey in four bands 3.4, 4.6, 12 \& 22 µm with  angular resolutions 6.1", 6.4", 6.5" \& 12.0", respectively. After the depletion of frozen hydrogen in October 2010, only two short wavelength bands (W1 and W2) can be used. NASA extended the mission by four months with a program called Near-Earth Object WISE (NEOWISE, \citealp{2011ApJ...743..156M}). The telescope was put into hibernation on 1 February 2011. On 21 August 2013, NASA announced that it would recommission NEOWISE (NEOWISE-R, \citealp{2014ApJ...792...30M}) to continue its search for near-Earth objects (NEO) and potentially dangerous asteroids. NEOWISE continued to observe from December 23, 2013.

The Wide-field Infrared Survey Explorer (WISE) scans the sky in large circles near a solar elongation of 90°, typically observing a given region of the sky over a period of one day every six months. unWISE co-adds group the exposures at each sky location into a series of six-monthly visits, with an average of typical 12 exposures per band per visit \citep{2014AJ....147..108L,2018AJ....156...69M}. Recently, \citet{2023AJ....165...36M} created the unTimely catalog, which contains sources extracted from time-resolved unWISE coadds. For each source detected in the unWISE coadd images, they provide parameters such as right ascension, declination, flux, flux error, fraction of flux in the source's point-spread function (PSF) that comes from the source, and local sky level. In this paper, we use the unTimely catalog (with 16 epochs) to identify variable sources.

\subsection{Select method} \label{Select method}
The focus of this paper is on sources with large-amplitude variability. Therefore, our objective is to identify sources with W1 band variability exceeding 1 mag, referred to as large-amplitude variables in the following text. Before beginning the process of selecting, we performed a cross-match between sources from different epochs within the designated area to create a WISE light curve catalog. This cross-matching was done using the $C^3$ (Command-line Catalog Cross-match) tool \citep{2017PASP..129b4005R} with a matching radius of 1".5. More discussion about the matching radius is given in the Appendix~\ref{Discussion on the crossmatch radius for Untimely data}. Our aim is to identify sources with changes of more than one magnitude. We achieved this by following three steps.

\begin{enumerate}
    \item Select reliable detections on a light curve with the following criteria: $fracflux>0.3$.This parameter represents the fraction of the flux in the image at the location of the source that is contributed by the source itself \citep{2018MNRAS.478..651G}.
    \item For the light curve consisting of reliable detections, we select sources that have a difference in magnitude of more than 1 magnitude. Alternatively, we consider sources where there was no detection in epoch-0 and epoch-1 and where the upper limits are less than 11 magnitudes. The latter criterion is appropriate in dense stellar fields. For more details, see Appendix~\ref{COMPLETENESS & RELIABILITY IN Dense stellar field}.
    \item Visually inspect all 16 epochs of unWISE coadds for potential artifacts and deblending problems.
\end{enumerate}

\begin{figure}
	\includegraphics[width=\columnwidth]{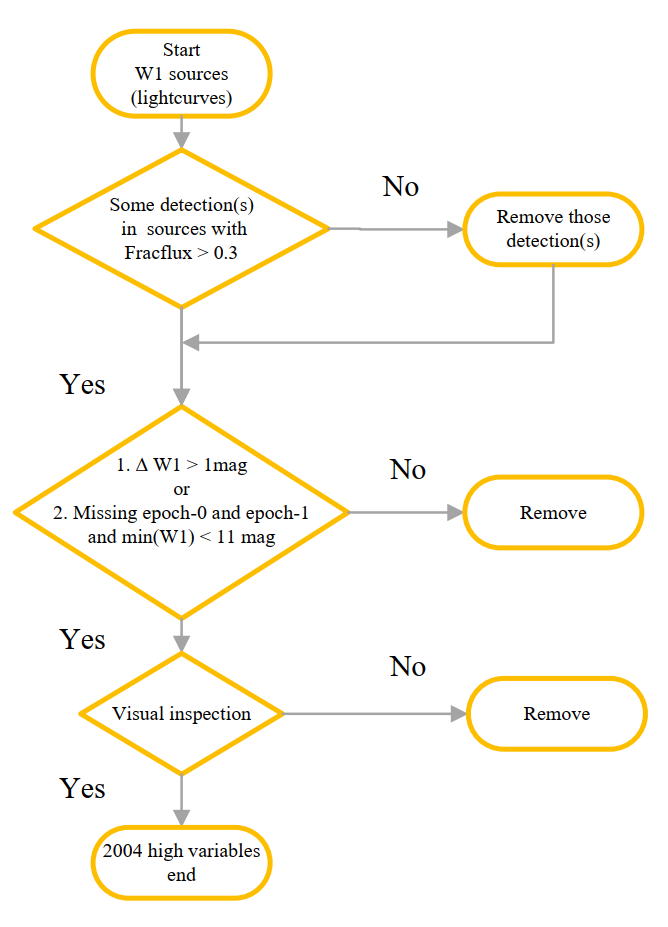}
    \caption{Summary flowchart of the variable selection process}
    \label{fig:flowchart1}
\end{figure}

The purpose of step 3 is to ensure the removal of artificial sources, which can include false detections or incorrect deblending, leading to incorrect photometry or misalignment of the photometry positions with the source center. We summarize our variable selection procedures in Figure~\ref{fig:flowchart1}. In the end, we identified a total of 2004 large-amplitude variables. We collected unTimely W2 data within a radius of 2" around the coordinates of these 2004 large-amplitude variables.

\subsection{Ancillary data}

In the subsequent analysis, we acquire multiband properties of sources in the sample by cross-matching in position with the following datasets. 

We utilized the VVV data from the VISTA (Visible and Infrared Survey Telescope for Astronomy) for our near-infrared analysis. 
The VVV program conducted multiband observations (ZYJHK$_\mathrm{s}$) with an exposure time of 80 s per filter at the beginning of the project. In 2015, a second set of contemporaneous (ZYJHK$_\mathrm{s}$) observations was carried out. For variability analysis, we focus exclusively on the K$_\mathrm{s}$ band data, which were obtained in a 16-second exposure time. In order to acquire NIR data for subsequent analysis, we matched these sources with the VVVDR5 detection table with 1" cross-matching radius. We also do some visual inspections (see Appendix~\ref{Further checking of the VVV data}). 

Photometric measurements from VVV were considered saturated if they met the following criteria: J<12, H<13, K$_\mathrm{s}$<11.5 \citep{2018MNRAS.474.5459G,2021ApJS..254...33K}. Additionally, we supplemented our analysis with data from the 2MASS All-Sky Point Source Catalog \citep{2006AJ....131.1163S} with 
1" cross-matching radius for three filters: J, H and K$_\mathrm{s}$.

For mid-infrared data, we utilized the GLIMPSE I archive \citep{2009yCat.2293....0S} from Spitzer \citep{2005AdSpR..36.1048W} and AllWISE \citep{2013wise.rept....1C} data. A matching radius of 2" is used for GLIMPSE I and AllWISE. GLIMPSE I comprises four mid-IR bands centered at 3.6, 4.5, 5.8 and 8.0 $\mu$m. The AllWISE catalog builds upon the achievements of the successful WISE by amalgamating data from both the WISE cryogenic phase and the NEOWISE post-cryogenic survey phases, offering the most comprehensive view of the complete mid-infrared sky currently available. AllWISE encompasses four mid-IR bands centered at 3.4, 4.6, 12 and 22 $\mu\mathrm{m}$.

\begin{figure}
	\includegraphics[width=\columnwidth]{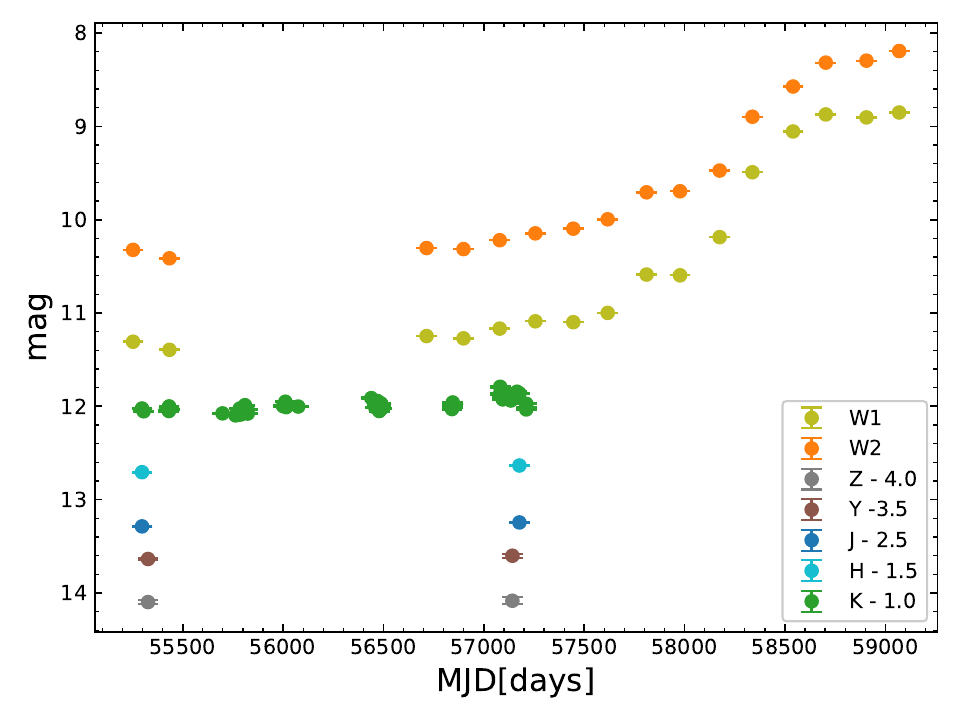}
    \caption{
    The light curve of one of the variable sources. Colors denote different band, grey for Z band, brown for Y band, blue for J band, cyan for H band, green for K$_\mathrm{s}$ band, yellowgreen for W1 band and orange for W2 band. ZYJHK$_\mathrm{s}$ are from VVV. W1 and W2 are from unTimely.
    }
    \label{fig:lightcurve}
\end{figure}

Figure~\ref{fig:lightcurve} presents the light curve of one variable source in our sample. AllWISE data and the early-stage data from VVV (mjd < 55500) were acquired at a similar time. In section \ref{YSO selection}, they will be used for the color-color selection of YSOs. The coordinates, light curve data, and multi-band data for these 2004 sources will be released online (see Appendix~\ref{multi-epoch photometry and sed data}).

\subsection{Completeness} \label{Completeness}
\citet{2017MNRAS.465.3011C} carried out a search for variable sources in the same region using the K$_\mathrm{s}$ band data from VVV. A total of 816 variable sources were identified, and only 747 of them were found within the designated research area. Among the latter, 189 sources were found to be common to our sample. In order to examine the 558 variable sources that were left, we compared the position of the sources with the unTimely W1 catalogue. It is worth noting that of these sources, 228 were not found in the unTimely W1 data in any epoch.

\begin{figure}
	\includegraphics[width=\columnwidth]{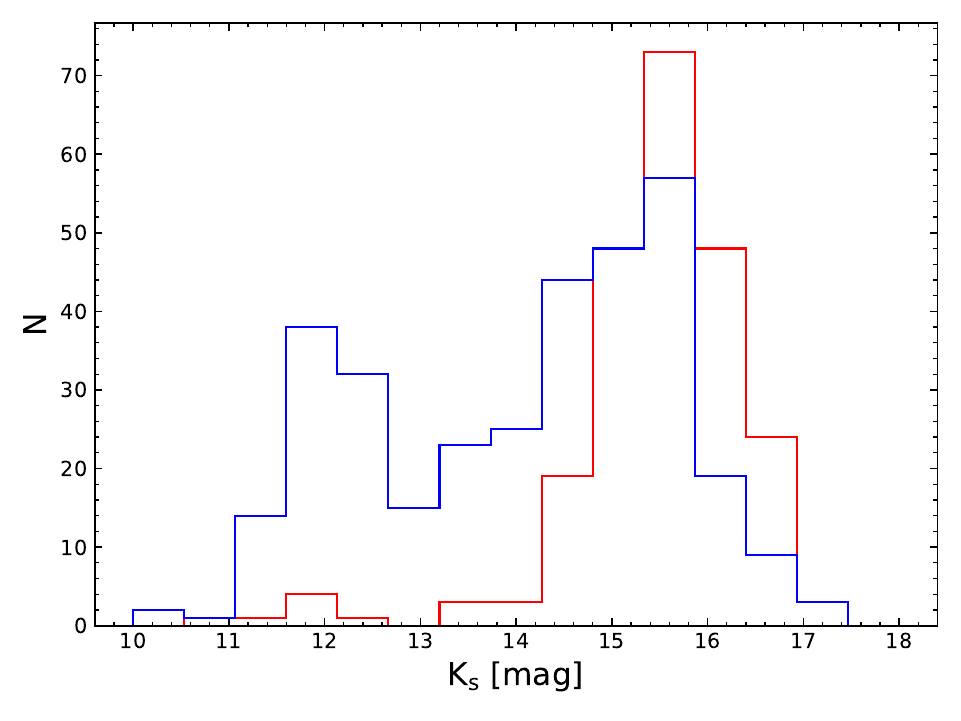}
    \caption{
    The K$_\mathrm{s}$ mag distribution of sources of 558 VVV variables which are not in our variable catalog. Red refers 228 sources did not have at least one W1 unTimely detection. Blue refers remaining 330 VVV variables. The K$_\mathrm{s}$ mag of each source are from \citet{2017MNRAS.465.3011C}
    }
    \label{fig:ks_df}
\end{figure}

These 228 sources have an average magnitude of 15.48 mag in K$_\mathrm{s}$, which is fainter than the average of 14.11 mag for the remaining 330 sources (see Figure~\ref{fig:ks_df}). 
As a result, we conclude that these 228 sources are weak MIR sources that fall below the detection limit of unTimely. 

Of the remaining 330 sources, 256 sources exhibited a $\Delta W1<1$ mag; 67 sources had a $\Delta W1>1$ mag, but did not pass our visual inspection due to bad deblending or artificial fake detection. Furthermore, 7 sources, undetected in both epoch-0 and epoch-1 with upper limits below 11 magnitudes, failed to meet our visual inspection criteria. Even if we consider all 67+7 sources as mid-infrared variable sources that we missed, our recall rate still remains at about 72.14\%. Therefore, within the detection limit of unTimely, most variable sources that meet the requirements of Section \ref{Select method} are found.

\section{YSO selection} \label{YSO selection}
To identify YSOs among the large-amplitude variables, we employed the following selection methods. Initially, we eliminated known non-YSOs by consulting the Simbad database \citep{2000A&AS..143....9W}. Next, we identified potential YSOs using the color-color method of \citet{2014ApJ...791..131K} (referred to as KL14 hereafter) or by cross-correlating with the SPICY database \citep{2021ApJS..254...33K}. Lastly, we eliminated any contamination from asymptotic giant branch stars (AGBs) and classical B-type emission-line stars (CBes) within the selected YSO candidates.

\subsection{Simbad}
We performed queries on the SIMBAD database to retrieve astronomical objects located within a radius of 2" around the YSO candidates. Subsequently, we excluded any known non-YSOs from our sample. These non-YSOs encompass a range of categories, including 23 AGBs, 2 Be stars, 1 blue supergiant, 3 carbon stars, 1 cataclysmic binary, 1 R CrB variable, 1 high-mass X-ray binary, 3 novae, 134 Miras, 60 OH/IR stars, 2 pulsating variabies and 1 symbiotic star.

\subsection{Color-color selection}
We obtained multi-band data for the large-amplitude variable sample through cross-matching in Section \ref{Mid-infrared variables selection}. To perform color-color selection, we used the average of VVV data with mjd $<$55500 for each VVV band. If a source was saturated in a specific band during the mjd $<$55500 period, the band photometry was masked and not included in the color calculation. For AllWISE data, W3 and W4 data in particular are vulnerable to error in star forming regions, owing to the use of list-driven photometry and the lack of a local background measurement, so we use the same criteria in section 3.2 of \citet{2014ApJ...791..131K} to remove the W3 and W4 bands that may be contaminated.
It is important to note that the VVV and AllWISE data were acquired in the same period, so we can derive MIR to NIR color from these data. 
This greatly simplifies the use of the color-color selection method for YSO identification.

We follow the approach of KL14 to identify YSOs. They outlined a YSO identification and classification scheme using AllWISE and 2MASS data, which they applied to photometry extracted from their own point source catalogs. In our study, we utilized near-infrared photometry from the VVV catalog instead of 2MASS because all our sources are large-amplitude variables. Furthermore, the VVV and AllWISE data were acquired in the same period.

The primary steps of the color-color selection process are shown in Figure 3 of KL14. Through this method, we identified a total of 655 YSOs.

\subsection{SPICY data}

On the Galactic plane, a dataset comprising 120,000 YSO candidates detected through Spitzer/IRAC was introduced by \citet{2021ApJS..254...33K}. The dataset also covers the sky region of ours.
The authors compiled their dataset by integrating IRAC photometry data from GLIMPSE and JHK$_\mathrm{s}$ data from 2MASS, UKIDSS, and VVV. To select YSO candidates, they employed a random forest classifier that was trained on confirmed YSOs. In our study, we conducted a cross-matching process between our variable sources and their catalog, which led to the identification of 731 YSOs. Of these, 433 YSOs had already been identified using the color-color selection method described above. We selected sources that met the color-color selection criteria or could be matched with SPICY, leading to a combined count of 953 YSOs.

\subsection{Remove AGB and CBes} \label{remove AGB}

Infrared color-magnitude or color-color diagrams often show overlap between AGBs, CBes, and YSOs because these objects frequently have cold disks or dense circumstellar material in common \citep{2013MNRAS.430.2910S,2021ApJ...916L..20L,2023ApJS..264...38M}. In our investigation, we used the two previously mentioned selection methods, based on color of sources. The SPICY catalog employed a random forest classifier, using color data as input \citep{2021ApJS..254...33K}. As a result, our selection may inadvertently include AGB and CBes objects due to potential misclassification.

Light curves have been shown to be effective in distinguishing between YSOs and AGBs. Pulsating AGBs, in particular, show magnitude variations that often follow distinct sinusoidal patterns \citep{2017MNRAS.465.3039C,2021ApJ...920..132P,2021ApJ...916L..20L}.

\begin{figure}
	\includegraphics[width=\columnwidth]{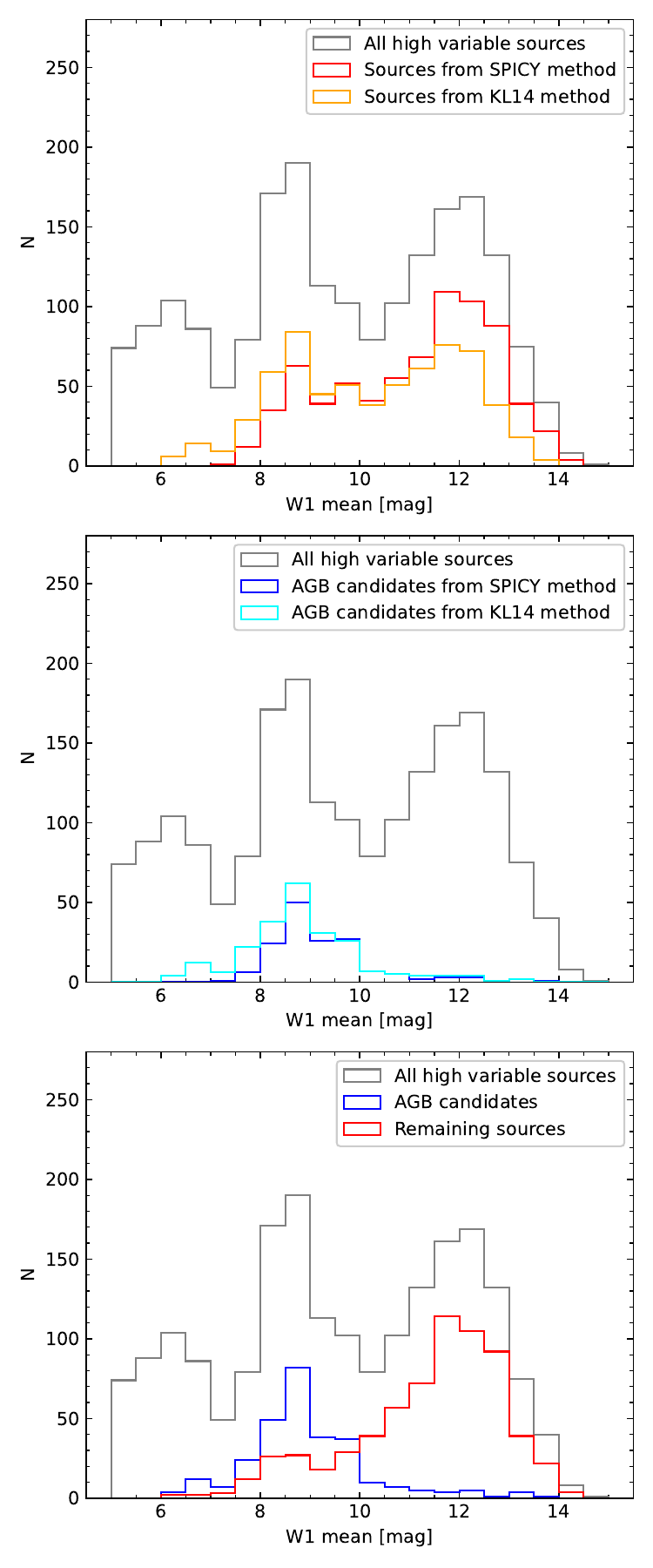}
    \caption{The distributions of mean W1 mag of variable sources. In the upper panel, grey, red and orange lines represent the entire variable sample, subsamples that selected with SPICY, KL14 methods, respectively. In middle panel, blue and cyan lines represent AGB candidates from SPICY method and KL14 method, respectively. The entire sample is also represented by gray line. In bottom panel, blue and red lines represent AGB candidates and remaining sources.}
    \label{fig:mag_df}
\end{figure}

We applied the Lomb-Scargle method to search for periodic variations in the magnitudes of W1 and W2. The Lomb-Scargle periodogram \citep{1976Ap&SS..39..447L,1982ApJ...263..835S,2018ApJS..236...16V} is a well-known algorithm for detecting and characterizing periodic signals in unevenly sampled data. According to \citet{2018ApJS..236...16V}, when we sample at a rate of $f_0=1/T$, we can fully recover the frequency information only if the signal's bandwidth is smaller than $f_0/2$, where $f_0/2$ is commonly referred to as the Nyquist frequency. For NEOWISE, with a sampling period of half a year, we set the minimum detectable period at one year. For the maximum detectable period, we set it to 5 years, which is half of the entire time baseline. We also introduce the concept of false alarm probability (FAP), which measures the likelihood that a data set without a signal would produce a peak of similar magnitude due to coincidental alignment among random errors \citep{2018ApJS..236...16V}. We obtained the FAP using the method proposed by \citet{2008MNRAS.385.1279B}. Subsequently, we eliminated the sources selected as mentioned above if FAP for W1 or W2 was less than $10^{-3}$, 
as these sources could potentially be AGB interlopers \citep{2020A&A...636A..48G}. 
Therefore, we remove 290 AGB candidates. Note that some short period AGBs may be missed due to the minimum period of one year.
The distribution of the infrared magnitudes of variable sources is presented in Figure~\ref{fig:mag_df}. It can be seen that before removing the AGB, both the YSOs identified through KL14 and those identified through SPICY exhibit a bimodal distribution in the magnitude distribution plot. The brighter peak in both cases corresponds to AGBs with periodic variations.
AGBs tend to spread in the brighter regime, a phenomenon also documented in research by \citet{2017MNRAS.465.3011C} and \citet{2013MNRAS.430.2910S}. Nonetheless, the removed AGBs may also contain some YSOs \citep{2022MNRAS.513.1015G}.

\begin{figure}
	\includegraphics[width=\columnwidth]{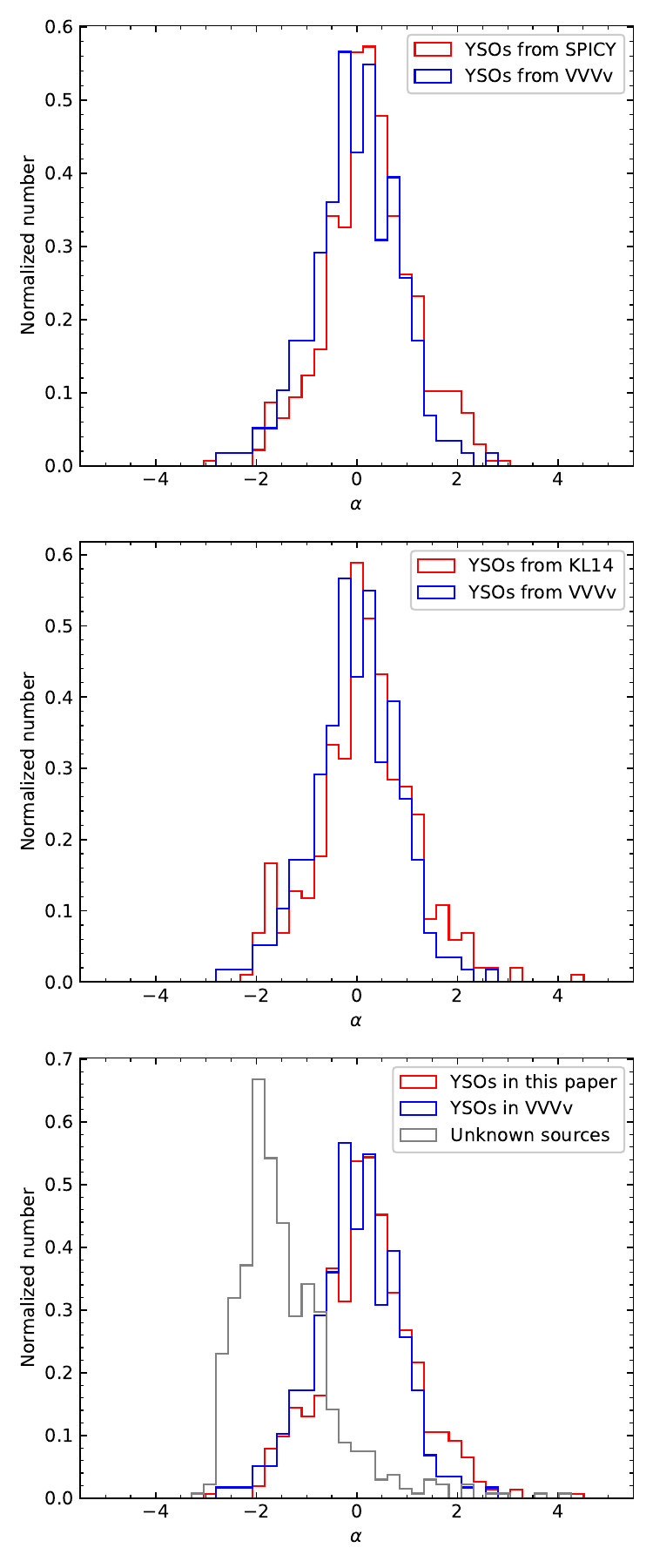}
    \caption{The distribution of Spectral index $\alpha$ defined in section \ref{Spectral index}. Blue refer 
 variable YSOs from \citet{2017MNRAS.465.3011C}. In upper panel, red refer our large-amplitude variable YSOs from SPICY. In middle panel, red refer our large-amplitude variable YSOs from KL14. In bottom panel, red refer our final large-amplitude variable YSOs, and grey refer unknown large-amplitude variables.}
    \label{fig:alpha_df}
\end{figure}

The spectral index of CBe sources generally falls within the range of -2.2 to -1.6 \citep[see][Figure 5]{2023ApJS..264...38M}. After excluding AGBs, we present the distribution of spectral indices for YSOs identified using SPICY and KL14 in Figure~\ref{fig:alpha_df}. Additionally, we display the spectral index distribution of young stellar variable sources selected by \citet{2017MNRAS.465.3011C}. The calculation method for the spectral index is described in Section \ref{Spectral index}. It is evident that the KL14 method yields a bimodal distribution in the spectral index, with a peak corresponding to the distribution of CBe sources. Conversely, the SPICY method exhibits a unimodal distribution, indicating a lower level of contamination from CBe sources. This suggests that the random forest method proposed by \citet{2021ApJS..254...33K} is more effective in removing CBes compared to the color-color selection based on KL14. Consequently, we eliminated the sources identified by KL14 within the spectral index range of -2.2 to -1.6, resulting in the removal of 22 CBe candidates.

\begin{figure}
   \includegraphics[width=\columnwidth]{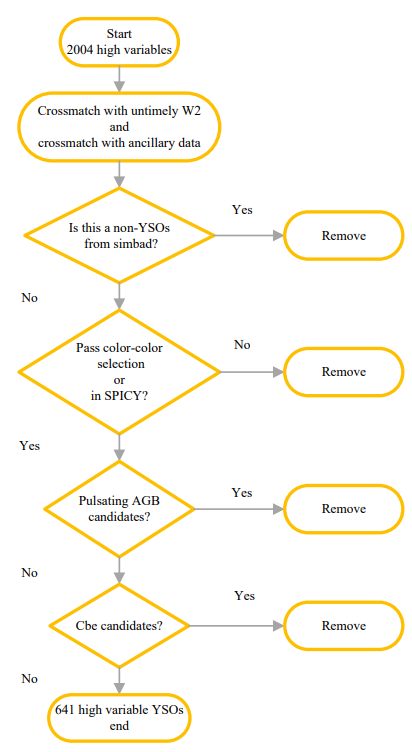}
    \caption{Summary flowchart of the YSO selection process}
    \label{fig:flowchart2}
\end{figure}

Before removing AGBs and CBes, among the 953 YSO candidates, 222 are not selected by SPICY and 298 are not selected by the KL14 method. After removing AGBs and CBes, of the remaining 641 sources, 68 are not selected by SPICY and 236 are not selected by the KL14 method. The intersection-to-union ratio is significantly improved. For the KL14 algorithm to function, it requires the presence of W1, W2, W3; or W1, W2, W3, W4; or H, K$_s$, W1. However, missing values often appear in W3 and W4. SPICY also certifies YSOs based on near-infrared and mid-infrared data, but for mid-infrared data, it uses Spitzer instead of AllWISE. SPICY, utilizing a missing value completion algorithm, can certify some YSOs that the KL14 algorithm cannot do due to its reliance on complete data sets. Consequently, after the exclusion of AGBs and CBes, there are typically more YSOs not certified by KL14 than by SPICY in the union process. Furthermore, SPICY benefits from the higher spatial resolution and better sensitivity of Spitzer compared to AllWISE.
Figure~\ref{fig:flowchart2} presents the flowchart used to identify YSOs. A total of 641 YSOs were discovered that exhibit large-amplitude variability. In Section \ref{Completeness}, a reassessment of 189 sources was conducted based on the variables identified by \citet{2017MNRAS.465.3011C}, which led to the discovery of 90 YSOs. Initially, \citet{2017MNRAS.465.3011C} had identified 87 YSOs. Compared with our sample, it was found that 75 YSOs were present in both studies. Therefore, the identification of YSOs in this study is considered to be reasonably complete and reliable.

\subsection{Spectral index} \label{Spectral index}

The spectral index is often used to describe the evolutionary phases of YSOs. We calculated the spectral index $\alpha$ following the method described by \label{1987IAUS..115....1L}: 
\begin{equation}
    \alpha = \frac{\mathrm{d} \log{(\lambda F_{\lambda})}}{\mathrm{d} \log{\lambda}},
\end{equation}
  Given the large-amplitude variations of the sources, it is of utmost importance to guarantee that an SED is created using photometry obtained from nearly simultaneous observations. In order to fulfill this requirement, the slope is derived from either Spitzer data or AllWISE data.

\begin{figure}
	\includegraphics[width=\columnwidth]{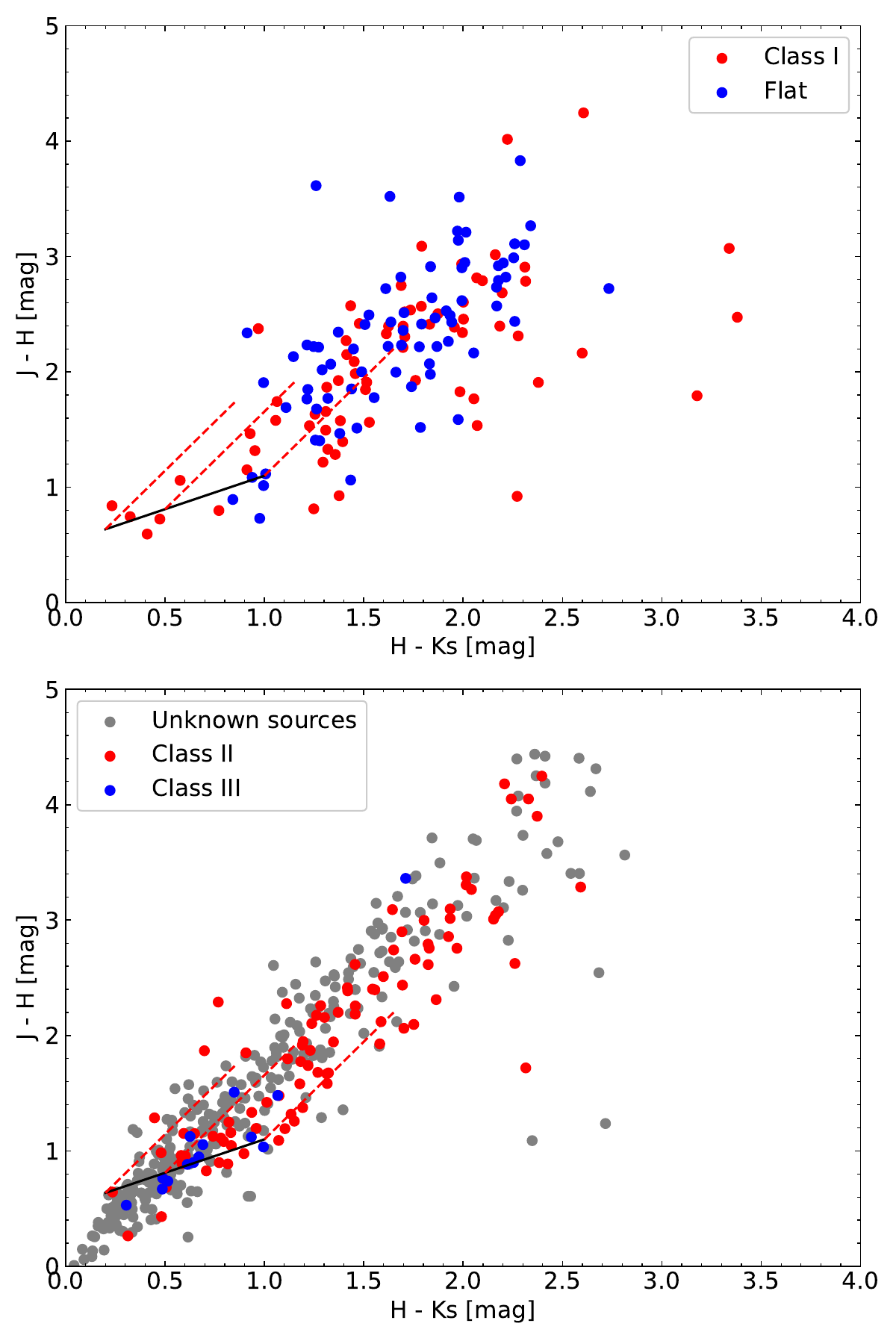}
    \caption{Top panel: The color-color diagram for class I YSOs (red) and flat YSOs (blue). Bottom panel: the diagram for Class II YSOs (red), Class III YSOs (blue) and unknown sources (grey). The solid black line represents the position of TTS (T Tauri star, \citealp{1997AJ....114..288M}), while the dotted red line represents the extinction line. The length of the black dotted line represents an extinction $A_V=10$ mag. }
    \label{fig:J-H-K}
\end{figure}

We perform a linear fit to the data from GLIMPSE I or AllWISE using scipy.optimize.curve\_fit. Where available, we prefer $\alpha$ based on GLIMPSE, followed by AllWISE. In the process of calculating the spectral index, we require that the source is detected at least in 3 bands. For calculations using GLIMPSE data, we also require detection in the I4 band. The four bands of GLIMPSE are simultaneous, as are those of AllWISE. With the spectral index $\alpha$, we can divide our selected YSO variables into the following four categories: Class I for $\alpha>0.3$, Flat for $-0.3\geq\alpha\leq0.3$, Class II for $-1.6<\alpha<-0.3$, Class III for $\alpha\leq-1.6$. Most of the YSO variables in our sample occupy the same region in the near-IR color space as typical YSOs. Figure~\ref{fig:J-H-K} illustrates the near-IR color-color diagram of the YSO variables and unknown variables for which JHK$_\mathrm{s}$ detections are available. The JHK$_\mathrm{s}$ values presented in Figure~\ref{fig:J-H-K} are obtained primarily from the average values in the VVV dataset with mjd < 55500 or from the 2MASS data. This is due to the saturation of one of the bands in JHK$_\mathrm{s}$ for certain sources, causing us to substitute all JHK$_\mathrm{s}$ values with those of 2MASS.

Furthermore, it becomes apparent that younger sources have a tendency to cluster towards the right side of the color-color diagram. This suggests that the classification outcomes derived from the mid-infrared spectral index align closely with those obtained from the near-infrared.

\section{Light curve classification} \label{Lightcurve classify}

\begin{figure*}
	\includegraphics[width=\textwidth]{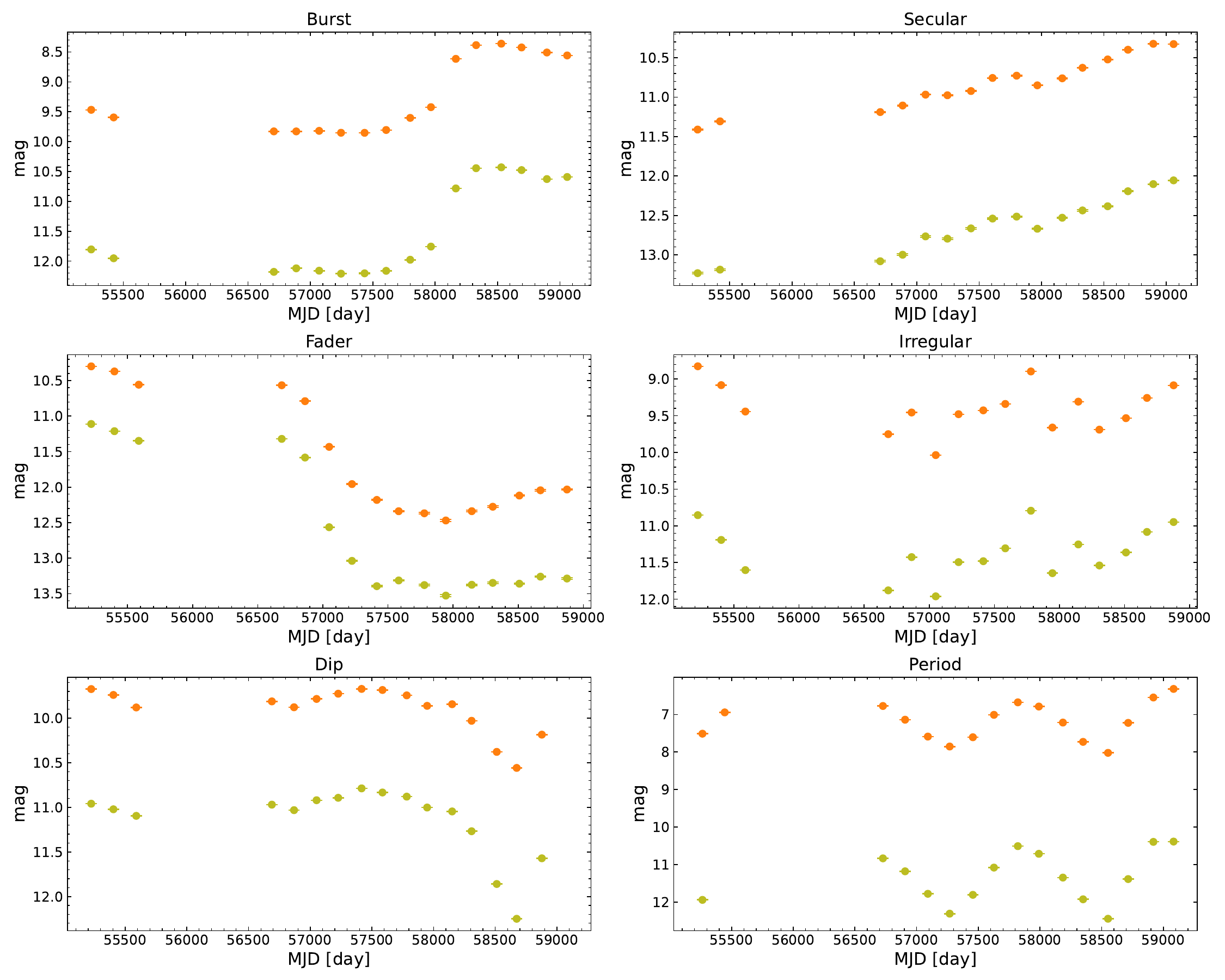}
    \caption{Examples of W1 (yellow-green) and W2 (orange) light curves for various types of large-amplitude variable sources. The left panels show examples of a burst, a fader, and a dip from the top to the bottom; while the right panels illustrate examples of secular, irregular, and period from top to bottom, respectively. 
    }
    \label{fig:6_examples}
\end{figure*}

\begin{figure}
	\includegraphics[width=\columnwidth]{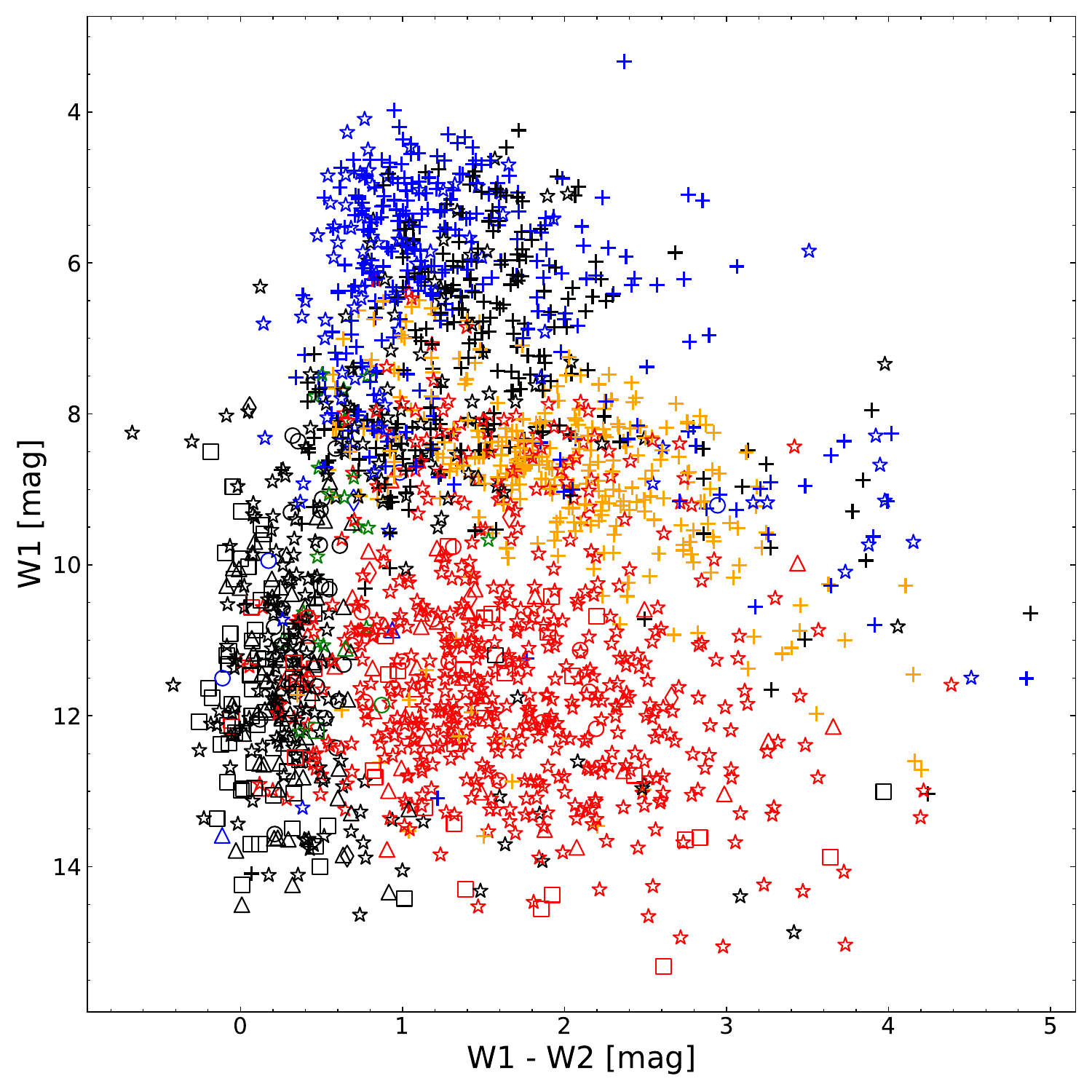}
    \caption{W1-W2 versus W1 for all large-amplitude variables. The different colors to represent different types. Red represents YSOs we selected. Blue represents non-YSO certified by simbad or KL14 method. Orange and green represent AGBs and CBes in section \ref{remove AGB} respectively. Black represents unknown sources. Different symbols to represent different types of light curves. The five-pointed star represents irregular sources. Triangles represent secular sources. The cross represents periods. Squares represent burst sources. The prism represents dip sources. The circles represent fader sources.}
    \label{fig:W1-W2_W1}
\end{figure}

We visually inspect the MIR light curves of large-amplitude variable YSOs. Based on their variability patterns, we classify them into the following categories: burst, fader, dip, secular, and irregular. Burst sources are characterized by a rapid increase in brightness in their light curve, which can occur once or multiple times. This increase in brightness is similar to that of FUors or EXors. The defining feature of fader sources is an extended decline in the light curve or a sudden drop, followed by a prolonged period of stability lasting at least one year. The dip sources exhibit sudden dimming events that occur once or multiple times, with subsequent recovery within a year. Among the remaining sources, those with smooth light curve changes are regarded as secular sources, while the other part of the sources is regarded as irregular sources. The rising and falling time scales corresponding to the former are much larger than the sampling interval of NEOWISE. The latter includes sources with comparatively chaotic and non-smooth light curves.
 Finally, we discovered 40 burst sources, 19 fader sources, 6 dip sources, 40 secular sources, and 536 irregular sources. It can be seen that the irregular class is predominant, consistent with the conclusions of \citet{2021ApJ...920..132P} and \citet{2023ApJS..264...38M}. 
In addition to classifying the light curves of the YSOs, we also classified the remaining 1363 sources. Among them, 290 periodic AGB candidates were found using the Lomb-Scargle periodogram method in Section 3.4. Excluding these 290 sources, we can find 526 periodic sources using the same method among the rest 1073 sources. Finally, the remaining sources are divided into the above five categories according to the visual inspection method of this program. In summary, we divide the 1363 sources that are not certified as YSOs into: 85 burst sources, 30 fader sources, 4 dip sources, 76 secular sources, 352 irregular sources and 816 periodic sources (including 290 AGB candidates).
Figure~\ref{fig:6_examples} shows examples of these six types. 
Table~\ref{tab:Statistics of high variables} provides information on these 2004 large-amplitude variables, including source index, coordinates, possible type, fitted spectral index, SED class, and type of light curve.
Figure~\ref{fig:W1-W2_W1} shows the color-magnitude diagram of these 2004 high-amplitude variables. It can be seen that the YSOs we selected are redder and fainter than other sources. The periodic sources are brighter than the other sources, which is consistent with Figure~\ref{fig:mag_df}. For unknown sources, the spectral index of these sources are smaller (see Figure~\ref{fig:alpha_df}), and the near-infrared and mid-infrared colors are bluer (see Figure~\ref{fig:J-H-K} and Figure~\ref{fig:W1-W2_W1}). We cannot completely rule out that they are YSOs, because there are a small number of YSOs with a small infrared excess or because they have experienced an burst, resulting in a relatively blue color. This article mainly discusses the previously selected YSOs that are redder in color. We also looked into the Gaia data and found that only 3\% YSOs in the sample have distance and luminosity information. While their location in the HR-diagram is consistent with YSOs, it give very limited information due to small numbers.

FUors and EXors typically exhibit variable amplitudes of 5-7 magnitudes in the optical wavelength range \citep{1996ARA&A..34..207H,2015ApJ...808...68H,2023ASPC..534..355F}. However, in the infrared, their amplitudes are only about two magnitudes \citep{2019ApJ...872..183F,2022RNAAS...6....6H}. FUors and EXors are two types of YSOs that exhibit different patterns of brightness variation over time. FUors start in a quiescent state and gradually increase in brightness over a period of months to years. Once they reach their peak brightness, they maintain this level for several years to even a century. On the other hand, EXors also start from a quiescent state but reach peak brightness much faster, typically within weeks to months. After reaching their peaks, the EXors then decline in brightness over a period of months to a year. In this study, our aim was to identify FUor candidates and EXor candidates in the burst type by looking for sources with amplitudes exceeding 2 magnitudes in the W1 or W2 bands. This threshold is reasonable in the mid-IR \citep{2022RNAAS...6....6H}. For FUor candidates, we specifically selected sources that exhibit typical light-curve patterns of FUors, characterized by a distinct rise followed by a plateau that persists for more than two years. However, there were two sources, V1298 and V1459, which were not initially detected in the W1 or W2 bands during the first two epochs due to their faintness. Nevertheless, based on their variability in the K-band, with an amplitude of approximately 4 magnitudes, these two sources are also considered as FUor candidates.
EXor candidates are selected on the basis of the following characteristics of light curves. These criteria involved identifying the distinct pattern of a sharp rise followed by a clear decline over a two-year period. In total, we found 18 FUor candidates and 1 EXor candidate in the burst dataset.

Among the 19 FUor/EXor candidates, 8 have been reported in previous studies. These sources are frequently linked to intricate structures like open clusters, bubbles, molecular clouds, and infrared dark-clouds (IRDCs). Therefore, we investigated the association of these sources with these established structures. \citet{2009AA,2019MNRAS.488.1141J,2021MNRAS.504..356D,2021MNRAS.500.3027D} provide lists of IRDCs, bubbles, open clusters and molecular clouds covering our study area, respectively. These lists contain the center coordinates and coverage areas of these structures. We checked whether these FUor(s)/EXor(s) are projected in the coverage of these structures. Note that for bubbles among them, YSOs often form at the edges of their structures. Therefore, when checking whether the bubbles were projected, we multiplied their geometric parameters (major semi-axis and minor semi-axis) by 1.5, and then checked. Finally, we found that 3 sources associate with bubbles, 4 sources associate with IRDCs and 2 sources associate with molecular cloud. All of this information is presented in Table~\ref{tab:FE information}. These light curves are shown in Appendix~\ref{The lightcurves of FUor/EXor candidates}.
All of them were listed in the SPICY catalog.

According to \citet{2023ASPC..534..355F}, it is not sufficient to rely solely on the characteristics of a star's light curve to classify it as a genuine FUor or EXor. The corresponding features must also be present in their spectra. Therefore, the subgroup of FUors/EXors identified on the basis of unTimely light curves can be considered as FUor/EXor candidates. For the sake of simplicity, we will refer to these candidates as FUor(s)/EXor(s) from now on.  

\begin{landscape}
\clearpage
\pagestyle{empty}
\setlength\LTleft{0pt}
\setlength\LTright{0pt}
\setlength\topmargin{-30pt}
\setlength\textwidth{702pt}
\setlength\textheight{38pc}
\begin{table*}
    \centering
    \caption{List of large-amplitude MIR variables. Only the first 10 sources are shown. The complete list is available in the online supplementary information.}
    \label{tab:Statistics of high variables}
    \begin{tabular}{|p{0.3cm}|p{1.2cm}|p{1.2cm}|p{0.8cm}|p{0.8cm}|p{1.2cm}|p{1.4cm}|p{1.2cm}|p{1.4cm}|p{1.0cm}|p{1.0cm}|p{1.5cm}|p{1.0cm}|p{1.5cm}|p{1.5cm}}
        \hline
        Index & RA & DEC & $ \Delta W1$ & $\Delta W2$ & Period(W1) & FAP(W1) & Period(W2) & FAP(W2) & KL14 & SPICY & Possible & $\alpha$ & class & light curve \\
         &  &  &  &  & (days) &  & (days)  &  & method & method & type$^\mathrm{a}$ &  & & classification\\
        \hline
        \begin{filecontents*}{type.csv}
Index,ra,dec,deltaw1mag,deltaw2mag,w1_period,w1_fap,w2_period,w2_fap,KL_method,SPICY_method,possible_type,alpha,class,lightcurve_type
V1,175.94801,-62.83249,1.02,1.18,--,--,--,--,--,YSO,YSO,-1.53,class II,irregular
V2,178.41579,-62.83414,1.29,1.61,--,--,--,--,--,--,unknown,--,--,secular
V3,178.21208,-62.84909,1.40,1.56,1659.22,$2.2\times10^{-1}$,1659.21,$1.3\times10^{-1}$,--,--,unknown,-2.41,--,burst
V4,175.67140,-62.82438,1.52,1.81,365.25,$9.0\times10^{-1}$,365.24,$9.1\times10^{-1}$,--,YSO,YSO,-1.50,class II,secular
V5,176.66073,-62.81652,2.42,2.17,1303.89,$6.5\times10^{-2}$,1303.88,$5.5\times10^{-2}$,YSO,YSO,YSO,-0.36,class II,fader
V6,177.77750,-62.72857,1.11,1.42,365.25,$2.2\times10^{-1}$,1404.11,$2.2\times10^{-1}$,--,--,unknown,-1.76,--,secular
V7,177.04208,-62.44676,1.28,1.41,--,--,--,--,YSO,--,CBe,-1.99,--,secular
V8,177.33415,-61.42612,2.36,2.19,480.57,$8.3\times10^{-4}$,480.55,$3.9\times10^{-4}$,--,--,Mira[simbad],0.13,--,period
V9,176.25583,-61.33202,1.12,1.06,--,--,--,--,--,YSO,YSO,0.59,class I,irregular
V10,175.78930,-62.35374,3.28,2.61,1521.05,$7.9\times10^{-1}$,1521.04,$6.5\times10^{-1}$,--,YSO,YSO,-0.25,flat,burst(FUor)
...,...,...,...,...,...,...,...,...,...,...,...,...,...,...
\end{filecontents*}
\csvreader{type.csv}{}%
        {\csvcoli & \csvcolii & \csvcoliii & \csvcoliv & \csvcolv & \csvcolvi & \csvcolvii & \csvcolviii & \csvcolix & \csvcolx & \csvcolxi & \csvcolxii & \csvcolxiii & \csvcolxiv & \csvcolxv \\}\\
        \hline
        \multicolumn{15}{l}{a. "Possible type refer to the classification results of these 2004 sources in section 3. The square brackets at the end of these categories are "[simbad]" and "[KL]",}\\
        \multicolumn{15}{l}{ which indicate that these non-YSO types are identified by only simbad or KL14 method respectively.}
    \end{tabular}
\end{table*}
\end{landscape}

\begin{table*}
    \centering
    \scriptsize
    \caption{The information of these 19 FUor(s)/EXor(s).}
    \label{tab:FE information}
    \begin{tabular}{|p{0.7cm}|p{1.3cm}|p{1.0cm}|p{1.0cm}|p{0.8cm}|p{3.2cm}|p{1.5cm}|p{1.5cm}|p{1.0cm}|p{0.5cm}|p{0.5cm}|}
        \hline
        Index & another name$^{\mathrm{a}}$ & RA & DEC & class & near object$^{\mathrm{b}}$ & lightcurve shape & spectrum feature & distance & $\Delta$ W1 & $\Delta$ W2  \\
         &  &  &  &  &  &  &  & (pc) & (mag) & (mag)\\
        \hline
        \begin{filecontents*}{FE_info.csv}
id,name,other name,ra,dec,spectra_index,near_object,lightcurve shape,spectrum feature,distance,deltaw1mag,deltaw2mag
54,V10$^*$$^\mathrm{c}$,L222\_1$^{[7]}$$^\mathrm{d}$,175.7893,-62.35374,Flat,Bubble[2G2950451-0056644]$^{[4]}$,FUor,FUor$^{[8]}$,10.2$^{+1.0}_{-0.8}$$^{[8]}$,3.28,2.61
232,V127,--,193.50537,-62.25289,Flat,--,EXor,--,--,2.31,2.75
260,V148$^*$,Stim 1$^{[9][6]}$,194.43414,-62.25175,Class I,--,FUor,EXor$^{[6]}$,--,2.1,1.64
284,V157,--,196.97276,-62.47722,Class I,--,FUor,--,--,3.12,2.01
335,V191$^*$,L222\_10$^{[7]}$,201.45941,-62.07971,--,--,FUor,FUor$^{[8]}$,3.7$^{+0.5}_{-1.0}$$^{[8]}$,4.04,3.09
525,V352,--,220.76843,-59.86017,Class I,--,FUor,--,--,2.65,2.37
659,V450,--,237.80828,-53.46381,Class II,--,FUor,--,--,2.54,2.22
861,V606,--,249.55999,-47.55991,Class II,IRDC[337.270-0.389]$^{[1]}$,FUor,--,--,3.07,2.79
907,V636$^*$,YSO3416$^{[9]}$,250.93629,-46.76075,Flat,--,FUor,--,--,2.07,2.19
1450,V978$^*$,L222\_18$^{[7]}$,214.07491,-61.37315,Flat,--,FUor,FUor$^{[8]}$,2.0$^{+0.5}_{-0.6}$$^{[8]}$,3.68,3.45
1961,V1298$^*$,VVVv270$^{[2]}$,245.86306,-49.74554,Flat,--,FUor,EXor$^{[3]}$,2.3$^{[3]}$,1.08$\uparrow$$^\mathrm{e}$,1.44
2045,V1349$^*$,VVVv721$^{[2]}$,249.95318,-45.81332,Class I,Bubble[2G3388931+0061032]$^{[4]}$,FUor,FUor$^{[3]}$,4.2$^{[3]}$,2.05,1.35
2217,V1459,--,252.57525,-44.73284,--,MC[SDG340.765-0.0853]$^{[5]}$,FUor,--,6.18$^{+0.45}_{-0.45}$$^{[5]}$,1.40$\uparrow$,0.53$\uparrow$
,,,,,,IRDC[340.769-0.119]$^{[1]}$,,,,,
2239,V1476,--,254.52377,-43.46259,Class II,--,FUor,--,--,2.12,1.15
2468,V1610,--,188.60311,-61.96189,Class I,Bubble[2G3009075+0088586]$^{[4]}$,FUor,--,--,1.46$\uparrow$,2.04
2487,V1625,--,208.26075,-61.42611,Class II,--,FUor,--,--,2.36,1.26
2497,V1633,--,224.08106,-58.32878,Class I,--,FUor,--,--,1.55$\uparrow$,3
2594,V1707$^*$,L222\_59$^{[7]}$,253.43492,-43.47208,Class I,MC[SDG342.136+0.2045]$^{[5]}$,FUor,Unusual$^{[8]}$,3.23$^{+0.51}_{-0.51}$$^{[8]}$,5.04$\uparrow$,3.23
,,,,,,IRDC[342.135+0.204]$^{[1]}$,,,,,
2795,V1823,--,204.83296,-62.33359,Flat,IRDC[308.458+0.025]$^{[1]}$,FUor,--,--,1.33$\uparrow$,2.4
\end{filecontents*}
\csvreader{FE_info.csv}{}%
        {\csvcolii & \csvcoliii & \csvcoliv & \csvcolv & \csvcolvi & \csvcolvii & \csvcolviii & \csvcolix & \csvcolx & \csvcolxi & \csvcolxii \\}\\
        \hline
        \multicolumn{11}{l}{a. This column indicates the names of these sources in other literature.}\\
        \multicolumn{11}{l}{b. This column represents the names of structures around these sources. See section \ref{Lightcurve classify}.}\\
        \multicolumn{11}{l}{c. The "*" at the end of the index of a source means that the source has been mentioned elsewhere in the literature.}\\
        \multicolumn{11}{l}{d. The superscript in square brackets indicates that the data were obtained elsewhere in the literature. The numbers 1-9 in square brackets represent documents}\\
        \multicolumn{11}{l}{\citet{2009AA,2017MNRAS.465.3011C,2017MNRAS.465.3039C,2019MNRAS.488.1141J,2021MNRAS.500.3027D,2021MNRAS.504..830G,2024MNRAS.528.1789L,2024MNRAS.528.1769G} and \citet{2024MNRAS.528.1823C}
        respectively.}\\
        \multicolumn{11}{l}{e.  The upward arrow indicates that the data is only the lower limit.}\\
    \end{tabular}
\end{table*}

\section{Result} \label{Result}
\subsection{SED class \& Light curve morphologies}
According to Section \ref{Spectral index}, we obtained the spectral indices for all large-amplitude variable YSOs, excluding those that had insufficient detections (at least 3 bands). In total, we identified 277 YSOs as Class I, 185 as Flat, 149 as Class II, 15 as Class III, and 15 with unknown SED class. The lower panel of Figure~\ref{fig:alpha_df} presents a comparison of the spectral index distributions between the YSOs in our sample and those of \citet{2017MNRAS.465.3011C}. The YSOs in our sample appear to be slightly younger compared to those studied by \citet{2017MNRAS.465.3011C}. 
We performed a KS-test between the spectral index distributions in their sample and in our sample.
 The test yielded a p-value of 0.0043, suggesting that the two distributions are statistically distinct. Specifically, the YSOs in our sample tend to exhibit a redder color.

In Section \ref{Lightcurve classify}, the variable YSO sources are classified into five different variability types. The lower two panels of Figure \ref{fig:alpha_colorline} present the distributions of the spectral indices for these types. Sources classified as dip and fader types exclude instances with high spectral indices ($\alpha > 1$). In contrast, the burst-type sources exhibit a more even distribution than the irregular and secular types. This suggests that bursts are more frequent among younger sources, or sources with a large mass of warm dust in their outer disk, compared to those in dips and faders. This finding is also supported by \citet{2017MNRAS.465.3011C}. 

\subsection{Variability amplitude with SED class} \label{Variability amplitude with SED class}
Typically, sources with higher spectral indices are associated with younger sources. This is because as YSOs evolve, their envelope tends to gradually vanish, followed by a gradual reduction and eventual disappearance of the disk. As a result, the radiation emitted by the envelope and disk decreases over time, leading to a decrease in the spectral index as stars age. It is important to note that factors such as the inclination of the disk or the temperature of the star can also affect the shape of the SED at the wavelengths used to determine the spectral index.

\begin{figure}
	\includegraphics[width=\columnwidth]{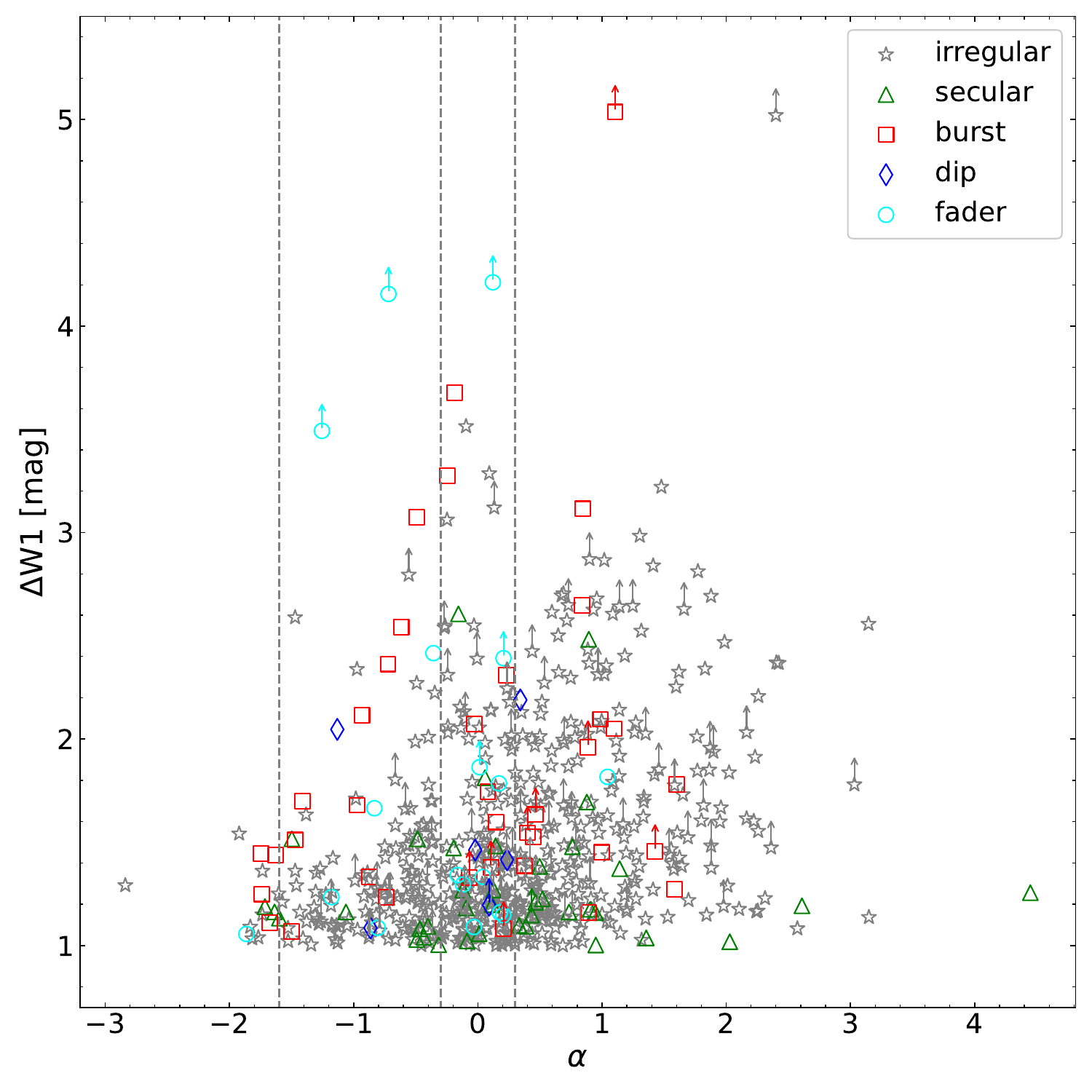}
    \caption{Variability amplitude $\Delta$W1 versus IR spectra index $\alpha$ for variable YSOs. Vertical dashed lines are used to indicate the boundaries between different classes of YSOs. Red, cyan, blue, green, and grey colors are used to represent burst, fader, dip, secular, and irregular sources, Where the arrow represents the lower limit.}
    \label{fig:w1_alpha}
\end{figure}

The variability amplitude in W1 ($\Delta$W1) versus the IR spectral index for various categories of light curves is depicted in Figure~\ref{fig:w1_alpha}. It is evident that there is a correlation between these two parameters, suggesting that a larger $\Delta$ W1 corresponds to a younger source. This result is consistent with the findings of \citet{2017MNRAS.465.3011C} and \citet{2021ApJ...920..132P}.

In Section \ref{Select method}, we performed a cross-match between the unTimely W1 light curve catalog and the SPICY catalog, which served as our parent sample. We retained the corresponding sources from the catalog of the unTimely W1 light curves, resulting in a total 27117 matches, of which 4675 are Class I, 5013 are Flat, 14155 are Class II, 1695 are Class III and 1639 are unknown SED class sources. However, as mentioned in Section \ref{Select method}, there may be false detection and incorrect deblending in the unTimely W1 light curves. Among these 27177 matches, we randomly selected 1000 sources to visually inspect their unTimely W1 images and found that 674 sources passed our visual inspection. In other words, about 67.4\% of the unTimely W1 light curves in these 27177 SPICY sources are valid. So, the number of valid SPICY sources in our region of interest is about 3151 Class I, 3379 Flat, 9540 Class II, 1142 Class III and 1105 unknown SED class sources, respectively.

Among these, we selected 573 variable YSO sources, comprising 256 Class I, 175 Flat, 119 Class II, 15 Class III, and 8 sources of unknown SED class. We assumed that the number of large-amplitude variable YSOs observed within a specific time frame follows a Poisson distribution. We choose a confidence level of 95\% to estimate the error of the proportions. 
Table~\ref{tab:class_bili} illustrates the proportion of variable sources within different YSO classes. In particular, younger groups exhibited a higher proportion of sources with variability. 

\begin{table}
	\centering
	\caption{Proportion of variable YSOs from different class}
	\label{tab:class_bili}
	\begin{tabular}{lccc} 
		\hline
		   & SPICY & large-amplitude & Proportion\\
              &       &  variable YSOs in SPICY &  \\
		\hline
		Class I & 3151 & 256$\pm$32 & 0.0812$\pm$0.0101\\
		Flat & 3379 & 175$\pm$26 & 0.0518$\pm$0.0077\\
		Class II & 9549 & 119$\pm$22 & 0.0125$\pm$0.0023\\
            Class III & 1142 & 15$\pm$8 & 0.0131$\pm$0.0070\\
            unknown & 1105 & 8$\pm$6 & 0.00724$\pm$0.0054\\
		\hline
	\end{tabular}
\end{table}

\begin{figure}
	\includegraphics[width=\columnwidth]{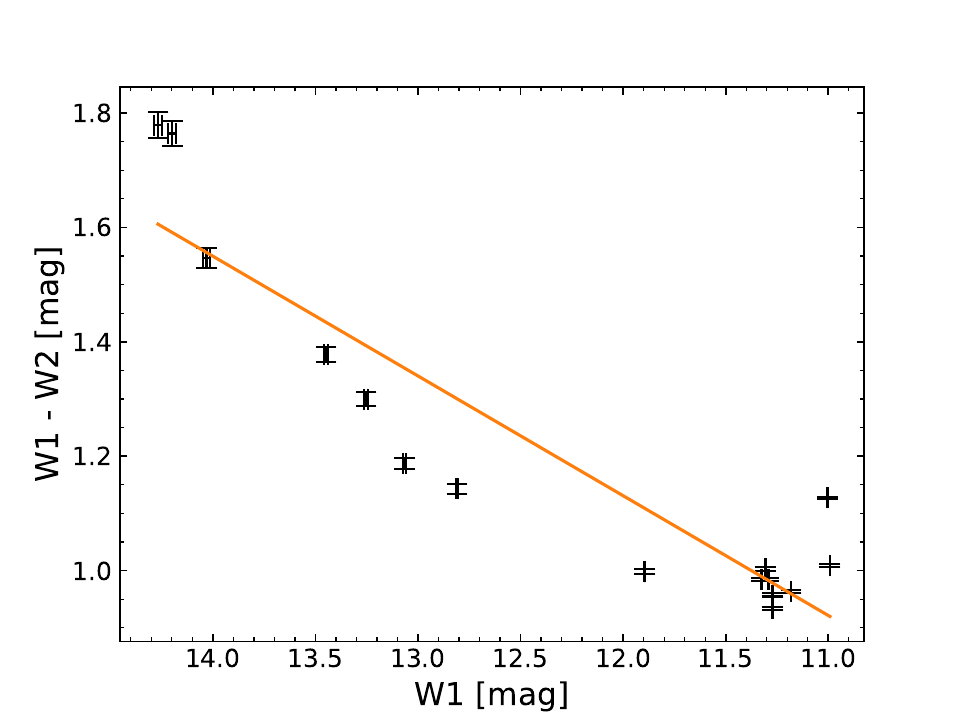}
    \caption{The color-magnitude relation for a YSO in the MIR band during periods of flux variations. The straight line represents the best fit obtained using the least squares method.}
    \label{fig:cc_example}
\end{figure}

\subsection{Mid-IR colour variability} \label{Mid-IR colour variability}
\begin{figure*}
	\includegraphics[width=\textwidth]{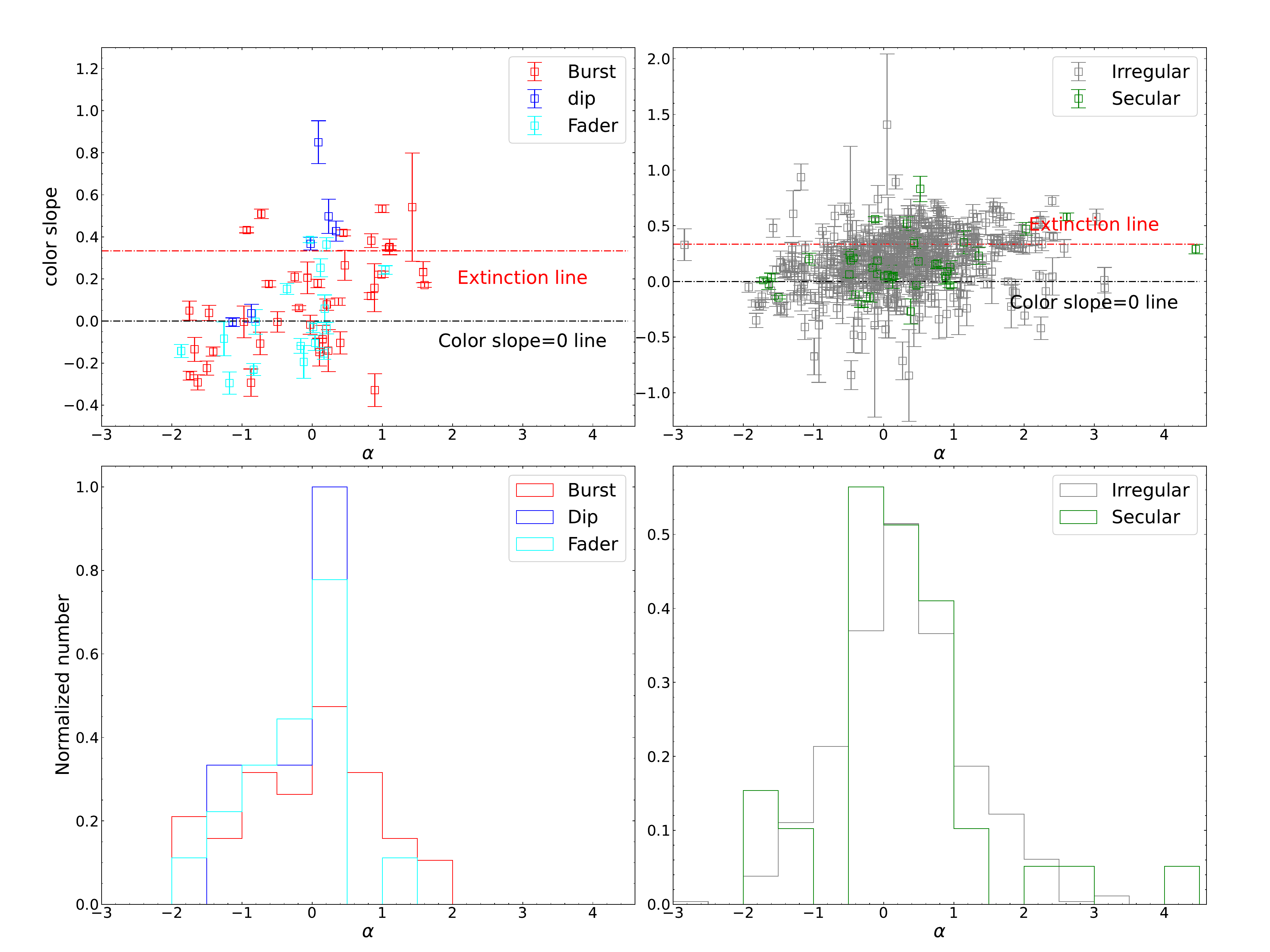}
    \caption{The upper two panel shows the spectra index $\alpha$ versus color slope of variable YSOs. 
     The color variation along the extinction direction is indicated by the red horizontal line color slope = 0.33, while the graph is divided into upper and lower sections by the blank horizontal line color slope = 0. The upper part indicates a pattern that is bluer when brighter, and the lower part represents a pattern that is redder when brighter. The lower two panels show the distribution of Spectral index $\alpha$ from bursts(red), dips(blue), faders(cyan), secular(green) and irregular(gray).
}
    \label{fig:alpha_colorline}
\end{figure*}

The color variations in YSOs are notably intricate, encompassing both "bluer when brighter" and "redder when brighter" trends, as well as more intricate alterations that defy straightforward categorization as "bluer" or "redder" with increasing luminosity. The majority of YSOs in the sample have unTimely data that are well matched for both W1 and W2 observations at the same time. As illustrated in Figure~\ref{fig:cc_example}, we constructed a color-magnitude diagram (CMD) depicting W1-W2 and W1 for each source and fitted a linear function to the CMD to obtain a color slope. In the upper two panels of Figure~\ref{fig:alpha_colorline}, the relationship between the color slope and the infrared spectral slope is shown for variable YSOs. It can be observed that among the five types of light curves, the variable YSOs include objects that become redder when brighter and those that become bluer when brighter. Most sources show no variability along the extinction curve, which is consistent with \citet{2017MNRAS.465.3011C}. 

We also calculated the Pearson correlation coefficient and its corresponding significance between the color slope and the spectral index for each type of light curve. 
Their correlation coefficients (P-values) are 0.51 ($1.1\times10^{-3}$), 0.81 ($4.9\times10^{-2}$), 0.53 ($2.9\times10^{-2}$), 0.40 ($1.3\times10^{-2}$) and 0.29 ($4.2\times10^{-11}$) for bursts, dips, faders, seculars and Irrgulars, respectively. 
If we choose the commonly used $P=0.05$ as the threshold for a significant correlation, there is a positive correlation between the color slope and the spectral index for all kinds. However, for dips, their p-values are very close to the threshold, which may be caused by the small number of sources. Whether they have this tendency is not clear. 
The correlations for bursts and faders are stronger than that for seculars and irregulars. In the case of burst and fader sources, it was observed that those with a lower spectral index ($\alpha < -1$), which suggests older sources, tend to display a pattern of mid-infrared variability that they become redder when brighter or bluer when fainter.

\subsection{The recurrence timescale of FUor outbursts} \label{The recurrence timescale of FUor outbursts}
The recurrence timescale of YSO outbursts was estimated using the methodology proposed by \citet{2013MNRAS.430.2910S} and \citet{2019ApJ...872..183F}. It is important to note that all FUors in the sample are included in the SPICY catalog, which is considered as the parent sample (see Section \ref{Variability amplitude with SED class} for more details). To briefly outline this methodology, if there are $N_p$ sources in the parent sample and $n_p$ sources experience eruptions within the observation window $\Delta t$, assuming that the duration from the onset of eruption to reaching a stable brightness is shorter than the observation time $\Delta t$, the recurrence timescale $I_b$ for each source is determined as:
\begin{equation}
    I_b = \frac{N_p\Delta t}{n_b}.
\end{equation}
To assess the uncertainty of the recurrence timescale $I_b$, we use the following approach. If we assume that $I_b$ is a known parameter, then the likelihood of observing $n_b$ FUors out of a sample of $N_p$ sources within the time window $\Delta t$ can be described by a binomial distribution. This can be mathematically represented as:
\begin{equation}
    P(n_b\vert I_b, N_p, \Delta t)=\frac{N_p!}{n_b!(N_p-n_b)!}(1 - e^{-\Delta t/I_b})^{n_b}(e^{-\Delta t/I_b})^{(N_p-n_b)}.
\end{equation}
We assume that the prior distribution of $I_b$ conforms to a uniform distribution for $I_b$ in the range of $10^2$ to $10^8$ years. This approach enables us to obtain the posterior distribution $P(I_b\vert n_b, N_p, \Delta t)$ using Bayesian statistics.

As explained in Section \ref{Lightcurve classify}, it is necessary to observe the FUors for a period of 2 years or more to determine whether a stable plateau phase is reached. This means that it is difficult to differentiate between FUor and EXor states if a source has experienced an outburst within the past 2 years. The dataset used for the observations covers a span of approximately 10 years, providing an 8-year baseline for detecting FUor outbursts. A total of 18 FUors were identified, all of which were detectable using SPICY.

\begin{figure}
	\includegraphics[width=\columnwidth]{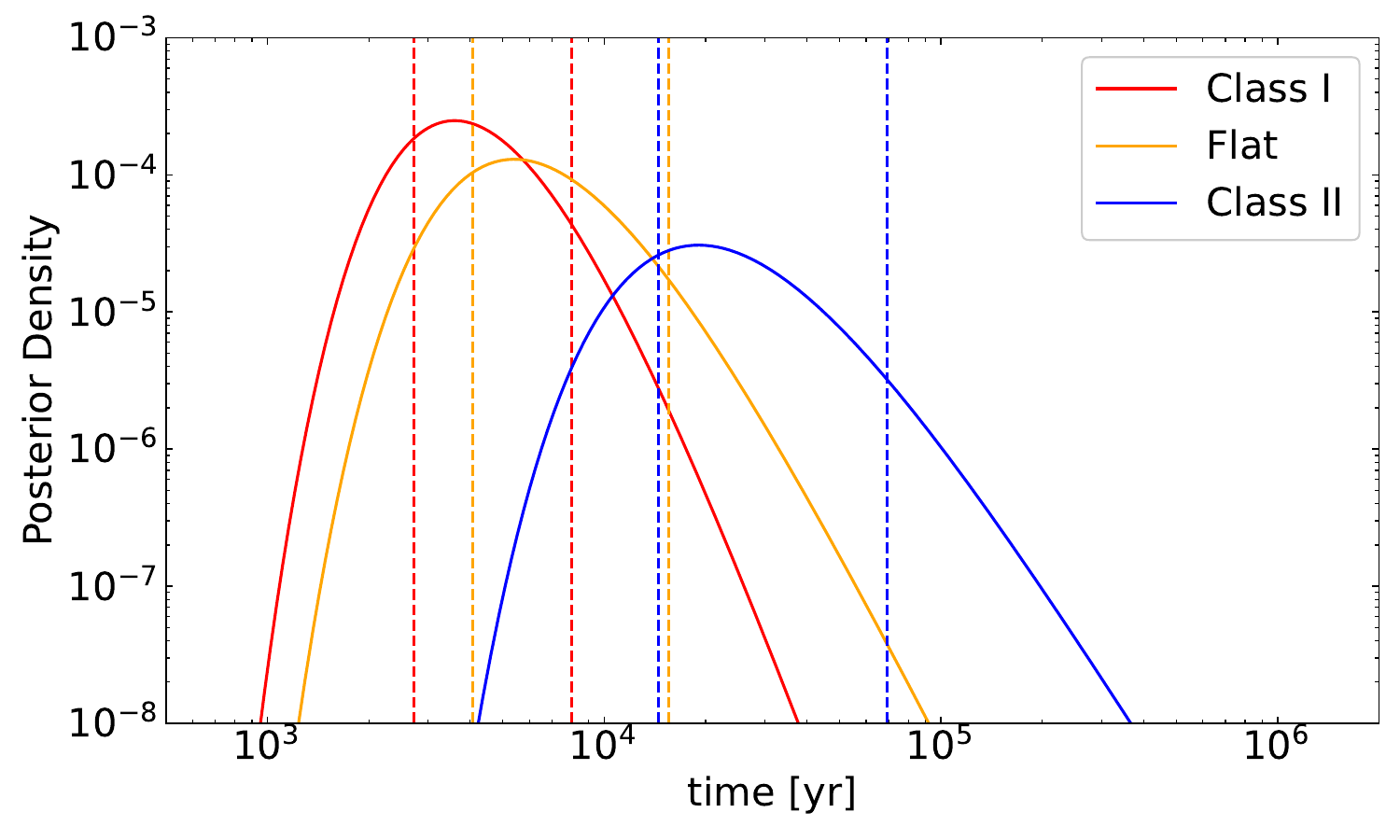}
    \caption{The $P(I_b\vert n_b, N_p, \Delta t)$ against $I_b$ for different class. Two vertical dotted lines of different colors represent the lower bound and upper bound of the 95\% confidence interval respectively. red for Class I, orange for flat, blue for Class II. We did not find FUor for Class III.}
    \label{fig:posterior}
\end{figure}

Figure~\ref{fig:posterior} illustrates $P(I_b\vert n_b, N_p, \Delta t)$ against $I_b$ for various classes. We used the 95\% confidence interval from the probability density function to establish upper and lower bounds. 
Note that the SED class for V191 and V1459 is somewhat uncertain due to the lack of I4 band in GLIMPSE and insufficient three bands in AllWISE. However, both of them have GLIMPSE data of three bands: I1, I2 and I3. If only the data of these three bands are used for calculation, their spectral indexes are 0.73 and 1.66 respectively, both belonging to Class I.
Additionally, no FUor instances were identified in Class III. Table~\ref{tab:recurrence_timescale_table} provides the recurrence timescales for FUor outbursts across different evolutionary classes.

Our analysis indicates that younger YSOs have shorter recurrence timescales. However, it is important to note that we only found one instance of EXor, whose spectral index belongs to a flat. This limited detection could be due to 
the sparse sampling frequency of WISE, which only provides observations every six months. It is worth mentioning that for most EXors, the decline phase usually starts several months after the initial outburst.

\begin{table}
	\centering
	\caption{Frequency of FUor Outbursts from different class}
	\label{tab:recurrence_timescale_table}
	\begin{tabular}{lccccc} 
		\hline
		   & SPICY & FUor & $\tau$ & $\tau_{\mathrm{lower}}$ & $\tau_{\mathrm{upper}}$\\
		\hline
		Class I & 3151 & 7 & 3593.81 & 2725.43& 8015.01\\
            Flat & 3379 & 5 & 5366.98 & 4070.14 & 15566.54\\
		Class II & 9540 & 4 & 19155.01&14526.54&69317.17\\
		\hline
	\end{tabular}
\end{table}

\section{Discussion} \label{Discussion}
\subsection{Comparison with others \& Mechanisms for YSO variability}

The observation period of this paper is up to 10 years, which is longer than most papers. In addition, the photometry of unTimely is from the unWISE images, which is deeper than directly from the NEOWISE images \citep{2023AJ....165...36M}. 
This will not only be more helpful and conducive to searching for long-time-scale variable sources with large amplitudes, such as FUors, but can also supplement sources with weak optical brightness among variable sources \citep{2020MNRAS.499.1805L,2023ApJ...957....8W}.
Furthermore, our focus is on sources that exhibit large-amplitude variations ($\Delta W1 > 1$ mag). However, the limited sampling of WISE over a period of only six months makes it challenging to capture the rapid changes in brightness accurately. In this study, we adopt a visual inspection-based approach for classifying light curves, which is similar to the method used by \citet{2017MNRAS.465.3011C}, rather than following the classification scheme proposed by \citet{2021ApJ...920..132P}. We have chosen not to follow the method proposed by \citet{2021ApJ...920..132P} because doing so would result in our sample's FUor candidates being categorized as either curved or linear, rather than burst-like. It is important to mention that nearly all FUor candidates identified by \citet{2021ApJ...920..132P} were classified as either curved or linear.

The study conducted by \citet{2011ApJ...733...50M} utilized data obtained from the YSOVAR project, which involved repetitive observation of the Orion nebula over a period of one month. Their results revealed that 70\% of the YSOs with disks and 44\% of the WTTs exhibited variability. Similarly, \citet{2014AJ....147...82C} examined YSOs in NGC 2264 using multi-band data, including the mid-infrared, for a period of one month. These authors discovered that a significant proportion of YSOs with disks, up to 91\%, showed infrared variability. In contrast, only about 36\% of the sources without disks displayed this variability. Another investigation by \citet{2018AJ....155...99W} used the YSOVAR project to track the mid-infrared variability of YSOs in the Serpens south for more than a month. They observed a high fraction of variable YSOs with disks, specifically 77\% ± 13\%, 74\% ± 20\%, and 59\% ± 19\% for SED classes I, flat, and II, respectively. Furthermore, \citet{2021ApJ...920..132P} analyzed 6.5-year NEOWISE-R data from known YSOs and found that the fractions of variables are 54. 8\%, 33.1\%, and 15. 4\% for Class 0 / I, flat, and II, III YSOs, respectively. 
Although our sample included only sources that varied by more than 1 magnitude over a duration of 10 years, it shows a similar trend that younger YSOs have a higher proportion of variable sources.

The threshold for our variable (1 mag) is identical to the one used by \citet{2017MNRAS.465.3011C}, the only difference being that they used the K$_\mathrm{s}$ band instead of the W1 band.
Compared to sources selected in the K-band \citet{2017MNRAS.465.3011C}, our variable sources are slightly redder.

Although we have classified light curves following \citet{2017MNRAS.465.3011C}, it is difficult to determine the physical mechanism of the light variation of these sources without spectra but only knowing the classification of light curves \citep{2021ApJ...920..132P}.
YSOs display a wide range of variations, from hours to months or even decades in terms of duration, and from a fraction of a magnitude to as high as 5-7 magnitudes in terms of the amplitude of optical variability. Variability in YSOs can be caused by a number of physical mechanisms, including changes in the accretion rate, occultation by an asymmetric structure on the accretion disk, or an increase in the height of the accretion disk, fluctuations in cold and hot spots on the star, and binary occultation.

For irregular variables, their variability occurs on a time scale that is shorter than the sampling interval of unWISE. Therefore, it is more challenging to determine their mechanisms through their light curves. 

In our study, we identified a total of six dips. Of these sources, two of them show a color slope that is almost zero, while the remaining four sources have a color slope that is equal to or greater than the extinction slope. The reason for the former is probably the shadowing caused by the optically thick material. The amplitude of its light change is only 1-2 mag in W1, which may be partially obscured by optically thick clumps.  
On the other hand, the higher color slope in the latter can be explained by the presence of a warped accretion disk that only obscures the bluer components of the interior or by the veiling during the occultation process \citep{2021AJ....161...61C}.

Bursts and faders exhibit large changes in brightness, followed by a period of stability or gradual change over several years. This pattern strongly suggests that these variations are likely caused by changes in accretion rates. Specifically, the burst corresponds to an increase in the rate of accretion, while the fader corresponds to a decrease. Nevertheless, it is worth noting that the observed light curves in some objects may also be attributed to occultation, as demonstrated in the case of AA Tau. In this case, the color slope should be equal to or greater than the slope of extinction \citep{2021AJ....161...61C}. However, the occurrence of such cases in bursts and faders is very low. 

The spectral index and color slope of these two types of sources show a positive correlation, with Pearson's correlation coefficients of 0.51 and 0.53, respectively. Specifically, sources with lower spectral index exhibit a redder when brighter (bluer when dimmer) tendency. On the other hand, sources with larger spectral indices showed an opposite trend of color variations (Figure~ \ref{fig:alpha_colorline}). 

The current YSO scenario can reconcile these patterns of variability. The observed infrared emission is a result of the combined contributions from the stellar photosphere, the inner accretion disk, and the reprocessing outer disk. When the accretion rates are low, the strong magnetic field in the YSO causes the inner disk to break, leading to matter falling onto the star along the magnetic field lines, a process known as column accretion. The majority of the accretion energy is released as UV and soft X-rays in the shocks above the stellar pole. In these situations, the disk's heating caused by viscosity is insignificant when compared to the heating from the star and accretion column, which is known as a passive disk. The YSO emission appears very blue in the infrared, while the reprocessed emission from the outer disk appears very red. As the accretion rate increases, the contribution of the passive disk becomes more significant because of the increased heating of the shocks. Moreover, the heating in the outer disk may also cause the disk to expand vertically, further enhancing the reprocessing component. 
Consequently, in low-accretion systems, the source becomes redder as the accretion rate increases.

However, this pattern will undergo a reversal when the accretion rate is high. As the accretion rate increases, the inner radius of the disk contracts, leading to an increasing significance of viscous heating. Eventually, the disk becomes the dominant source of infrared emission from the YSO. As the accretion rate increases, so does the temperature of the disk, resulting in a bluer infrared color. This phenomenon elucidates the observed trend of being bluer when brighter in redder systems \citep{2007ApJ...669..483Z,2022ApJ...936..152L,2022ApJ...927..144R}. 

For secular objects, their variability occurs on longer timescales than the sampling interval of unWISE, and the magnitude of their variability is smaller compared to bursts and faders, as shown in Figure~\ref{fig:w1_alpha}. The causes of their variability could be linked to occultation by an asymmetrical structure on the disk or fluctuations in accretion rates. However, it is improbable that short-term changes in light, such as star flares and hotspots, are responsible for this variability.

\subsection{Previous recurrence time scale of YSOs of outburst}

In recent years, there has been an increase in the availability of infrared data, prompting some researchers to focus on determining the frequency of EXor/FUor eruptions. \citet{2013MNRAS.430.2910S} utilized two separate sets of infrared data obtained from the Spitzer and WISE telescopes, which were taken 5 years apart. This allowed them to identify sources that experienced outbursts within a well-established sample of 23,330 YSOs. The selection criteria involved a significant difference in magnitude, exceeding 1 magnitude, in both the 3.6 $\mathrm{\mu}$m and 4.5 $\mathrm{\mu}$m bands. By detecting a limited number of YSOs that undergo outbursts, they were able to narrow the recurrence timescale for such eruptions to a range of 5000 to 50000 years. \citet{2019ApJ...872..183F} adopted a similar approach, with the inclusion of the 24 $\mathrm{\mu}$m band, to identify YSO outbursts within Orion molecular clouds. Over a period of 6.5 years, these authors were able to identify two outbursts in a group of 319 protostars. This suggests that protostars experience outbursts approximately every 1000 years, with a 90\% confidence interval ranging from 670 to 40300 years. The study of \citet{2019ApJ...883....6U} focused on massive protostars. Using NEOWISE time-domain data, they successfully identified five variable YSOs in a sample of 809 massive protostars. \citet{2021ApJ...920..132P} compiled W2 light curves for a total of 5398 known YSOs and subsequently identified 1734 variable sources. These variable sources were divided into six categories: linear, curved, periodic, burst, drop, and irregular. Furthermore, these authors visually examined the long-term light curves and identified FUor candidates from these sources. They provided an estimate of the recurrence time of the FUors, which ranged from hundreds to thousands of years.
Their approach is comparable to ours; however, the estimated recurrence time appears to be shorter. The reason may be that they incorrectly counted several known FUors (e.g., HOPS383) that
eruptions took placed before the NEOWISE surveys.
However, it is of utmost importance, as stressed in Section \ref{The recurrence timescale of FUor outbursts}, that the estimated timing of source bursts utilized for temporal scaling remains within the timeframe of observation.
\citet{2022ApJ...924L..23Z} conducted a comprehensive analysis to identify outbursts originating from Class 0 protostars in the Orion region. They used Spitzer photometric data, covering a time span of 13 years. Through their analysis, they were able to detect a total of five outburst events, of which three were classified as Class 0 outbursts. Based on these findings, it can be inferred that class 0 protostars experience bursts approximately once every 438 years, with a confidence interval 95\% ranging from 161 to 1884 years. In a recent study, \citet{2024MNRAS.528.1823C} investigated variability among approximately 7000 Class I YSOs in the K$_\mathrm{s}$ band over a period of 19 years. They discovered 97 eruptive variable YSOs, with 43 of these identified as long-term variables, and calculated a recurrence timescale of 1.75 kyr for Class I YSOs. Our findings are longer, likely due to differences in the bands used for the search.

In this paper, we focus on the region defined by $295^\circ<l<350^\circ$ and $-1^\circ<b<1^\circ$. Compared to previous studies, we use the unTimely catalog, which grants us access to data boasting a superior signal-to-noise ratio compared to the NEOWISE photometric data and a more extensive observational baseline, comprising approximately 16 epochs over 10 years. 
The expanded time period and increased number of data points in the unTimely WISE catalog improve our ability to effectively characterize FUors compared to having only two observations. It is important to clarify that, in this case, the term FUors refers to potential FUor candidates rather than those confirmed by spectroscopic analysis. The recurrence timescale of the FUors obtained in our study aligns with the findings reported in the cited article. We found only one EXor, which could be a conservative approximation.

\section{Conclusions} \label{Conclusions}
We conducted a comprehensive search for sources that show large-amplitude variations in brightness of more than 1 magnitude in the W1 band within the region of $295^\circ<l<350^\circ$ and $-1^\circ<b<1^\circ$. Subsequently, we used color-color diagrams and cross-matching with SPICY catalogs to identify YSOs among these sources. The initial sample was found to have some contamination by AGBs and CBes. The contamination from AGBs can be mitigated by removing periodic MIR variables.
Suspicious CBes were removed on the basis of the spectral index. As a result of our procedure, we successfully identified a total of 2004 large-amplitude variable sources, of which 641 are candidate variable YSOs.

Using GLIMPSE I and AllWISE data, we were able to categorize the YSOs into four different classes: 277 Class I, 185 Flat, 149 Class II and 15 Class III. There were also 15 sources that could not be classified due to unknown spectral indices. We found that the YSOs identified in this study appeared to be slightly younger compared to those in \citet{2017MNRAS.465.3039C}. When we considered SPICY as the parent sample, we observed a decrease in the fraction of large-amplitude variable YSOs as they evolved. Furthermore, we found that younger or embedded sources tend to have a larger and broader range of variability amplitudes.

In addition, we visually inspected the light curves of the large-amplitude variable YSOs using the unTimely data. We categorized these sources on the basis of their variability patterns, namely: burst, fader, dip, secular, and irregular. Irregulars dominate among them. Our analysis revealed that bursts are relatively younger compared to dips and faders. Furthermore, both bursts and faders exhibit larger variability amplitudes. To further investigate the behavior of these sources, we examined changes in W1-W2 versus W1 over a 10-year observation period. Intriguingly, we found that for sources classified as bursts and faders, with a lower spectral index $\alpha<-1$, tend to exhibit a MIR variability pattern of redder when brighter (or bluer when fainter). 
Interestingly, we found that for bursts and faders, sources with a lower spectral index ($\alpha < -1$) tend to exhibit a reddening of mid-infrared variability when brighter (or a bluing when fainter). These trends can be understood in the context of current accretion scenarios \citep{2022ApJ...936..152L}.

Among the burst types, we successfully identified a total of 18 FUor candidates and 1 Exor candidate. Moreover, using SPICY as the parent sample, we found that younger YSOs have a higher proportion of variable sources. We also estimate the recurrence timescales of the FUor candidates in different classes. Specifically, for Class I, Flat, and Class II, the estimated recurrence timescales are $3593.81_{-868.38}^{+4421.20}$ years, $5366.98_{-1296.84}^{+10199.56}$ years, and $19155.01_{-4628.47}^{+50162.16}$ years, respectively. It should be noted that no FUor was detected within class III.

\section*{Acknowledgements}

This research is supported by NSFC funding (NSFC-11833007). We thank the referee Dr Lucas for very constructive and helpful comments. In addition, this research has used the NASA/IPAC Infrared Science Archive, which is funded by the National Aeronautics and Space Administration and operated by the California Institute of Technology.

This publication makes use of data products from the Wide-field Infrared Survey Explorer, which is a joint project of the University of California, Los Angeles, and the Jet Propulsion Laboratory/California Institute of Technology, funded by the National Aeronautics and Space Administration. This publication also makes use of data products from NEOWISE, which is a project of the Jet Propulsion Laboratory/California Institute of Technology, funded by the Planetary Science Division of the National Aeronautics and Space Administration.

Data from VVV and 2MASS were used in this paper. We acknowledge Nick Cross from VSA-support for helping me with the VVV data. 2MASS is a joint project of the University of Massachusetts and the Infrared Processing and Analysis Center/California Institute of Technology, funded by the National Aeronautics and Space Administration and the National Science Foundation. We also acknowledge Zheyu Lin for suggestions on the format and style of the figures in this paper. 

\section*{Data Availability}
The unTimely data underlying this article are publicly available at the \href{https://portal.nersc.gov/project/cosmo/data/unwise/neo7/untimely-catalog}{https://portal.nersc.gov/project/cosmo/data/unwise/neo7/untimely-catalog}. The 2mass, Glimpse and AllWISE data are all publicly available at \href{https://irsa.ipac.caltech.edu/frontpage}{https://irsa.ipac.caltech.edu/frontpage}. The VVV data are also publicly at \href{http://vsa.roe.ac.uk}{http://vsa.roe.ac.uk}.




\bibliographystyle{mnras}
\bibliography{bib} 




\appendix
\section{On the cross-match radius for unTimely data} \label{Discussion on the crossmatch radius for Untimely data}
We have chosen two time periods, namely epoch-0 and epoch-2, within a specific region of interest (coadd\_id=2571m409). In epoch-0, we randomly selected 1000 sources and then determined the average number of sources that each of these 1000 sources could be associated with at various distances in epoch-2. While we ideally expect all 1000 sources from epoch-0 to be present in epoch-2, this is not realistic due to the high density of stars in the studied region, which may result in some false detections. Hence, we introduce the probability $\alpha$ to represent the likelihood of finding a matching source in epoch-2 for a given source in epoch-0. As the cross-match radius increases, more sources will be matched, although these sources may be neighboring sources rather than the exact counterparts of the sources in epoch-0.

\begin{figure}
    \centering
        \includegraphics[width=\columnwidth]{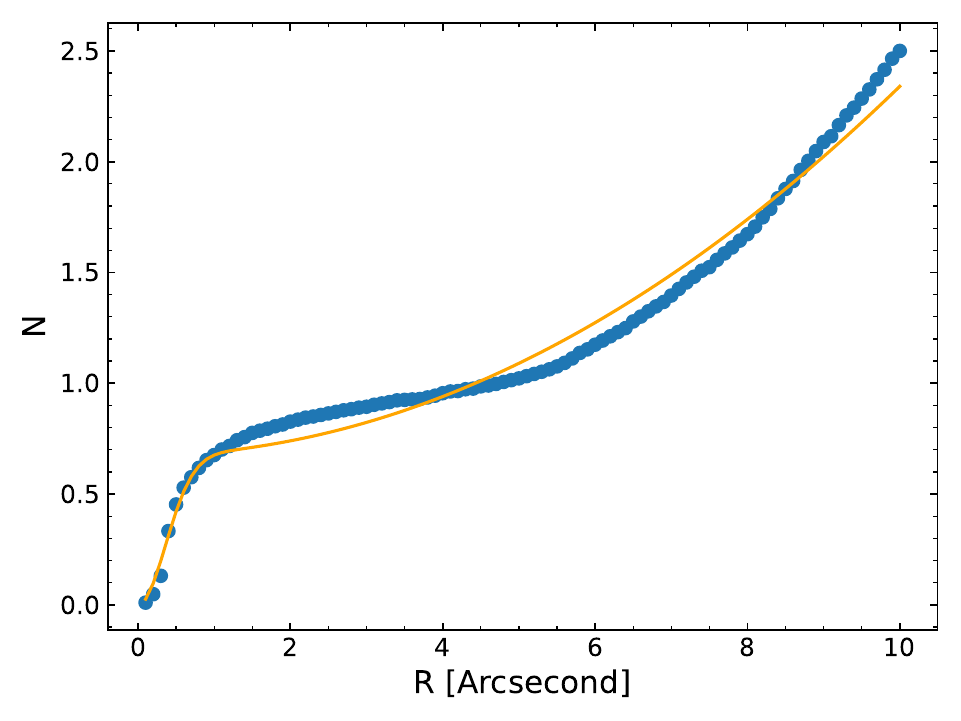}
    \caption{The average number of matches for the 1000 sources selected in epoch-0 at different radius (blue dot) in epoch-2 and the curve fitted through the equation\eqref{eq:mathematical expectation} (orange curve).}
    \label{fig:r_vs_N}
\end{figure}

For the sources in epoch-2 that correspond to the selected 1000 sources in epoch-0, we assume their coordinate distribution should follow a two-dimensional normal distribution centered at these sources in epoch-0. Let us denote the standard deviation of this distribution as $\sigma$. The unit for the standard deviation $\sigma$ is in arcseconds. We assume that the remaining sources in epoch-2 should follow a uniform distribution. Let us assume there are $\beta$ sources per square arcsecond on average. Therefore, for the sources selected in epoch-0, the number of non-self-matching sources within a certain area around them should follow a Poisson distribution with a parameter $\lambda = \pi R^2 \beta$. Let $X$ be a random variable representing the number of sources within a radius $R$ in epoch-2 for a particular source in epoch-0. The probability of matching to $k$ sources for each source selected in epoch-0 is given by:
\begin{equation}
\begin{aligned}
    p(X=k)=&\alpha \int_0^R \frac{r}{\sigma^2} \mathrm{e}^{-\frac{r^2}{2\sigma^2}} \mathrm{d}r \times \frac{\mathrm{e}^{-\pi R^2 \beta}(\pi R^2 \beta)^{k-1}}{(k-1)!}\\&+(1-\alpha\int_0^R\frac{r}{\sigma^2}\mathrm{e}^{-\frac{r^2}{2\sigma^2}} \mathrm{d}r)\times \frac{\mathrm{e}^{-\pi R^2 \beta}(\pi R^2 \beta)^{k}}{k!}
\end{aligned}
\end{equation}
where the integrand is the Rayleigh distribution. The first term on the right side of the equation represents that there are a self-match and $k-1$ non-self matches within the $R$ radius, and the second term on the right side of the equation represents that there are no self-match and $k$ non-self matches within the $R$ radius. The $k$ matches within the $R$ radius only include the above two situations. After integrating, its mathematical expectation is:
\begin{equation} \label{eq:mathematical expectation}
    E(X)=\alpha (1-\mathrm{e}^{-\frac{R^2}{2\sigma^2}})+\beta \pi R^2
\end{equation}
The blue dots on Figure~\ref{fig:r_vs_N} represent the average number of matches for each source in epoch-0, considering different radii in epoch-2. The orange curve is a fitted curve using the equation \eqref{eq:mathematical expectation}. 
The result of our fitting is: $\alpha=0.673$, $\beta=0.00530$ and $\sigma=0.360$. 

In this study, we have selected a radius of 1.5 arcseconds, which is approximately equivalent to 4 sigma. If there is a corresponding source ($\alpha$=1), the probability of this source falling within the specified region is 0.9998. It is evident that as the radius increases, the probability of a source falling within it also increases. However, when using a larger radius, there is a possibility of nearby sources falling within the region. Since our matching rule is to match the nearest source within 1.5 arcseconds, we need to determine the probability that the source corresponds to itself, given that there is a source within this radius. Let event A denote the nearest source in epoch-2 corresponding to itself, and event B represent the presence of sources within the radius $R$.
\begin{equation}
    P(A\cap B)=\int_0^R\frac{\alpha r}{\sigma^2}\mathrm{e}^{-\frac{r^2}{2\sigma^2}}\mathrm{e}^{-\pi r^2 \beta} \mathrm{d}r
\end{equation}
\begin{equation}
    P(B)=1-(1-\alpha\int_0^R\frac{r}{\sigma^2}\mathrm{e}^{-\frac{r^2}{2\sigma^2}}\mathrm{d}r)\mathrm{e}^{-\pi r^2 \beta}
\end{equation}
So, according to Bayes' theorem, we can calculate that when $R$=1.5.
\begin{equation}
    P(A|B)\Bigg|_{R=1.5}=\frac{P(A\cap B)}{P(B)}\Bigg|_{R=1.5}=0.9782
\end{equation}
Therefore, further visual inspection is needed to eliminate errors in variability detection caused by matching nearby sources.

\section{COMPLETENESS \& RELIABILITY IN Dense stellar field} \label{COMPLETENESS & RELIABILITY IN Dense stellar field}

\begin{figure}
    \centering
        \includegraphics[width=\columnwidth]{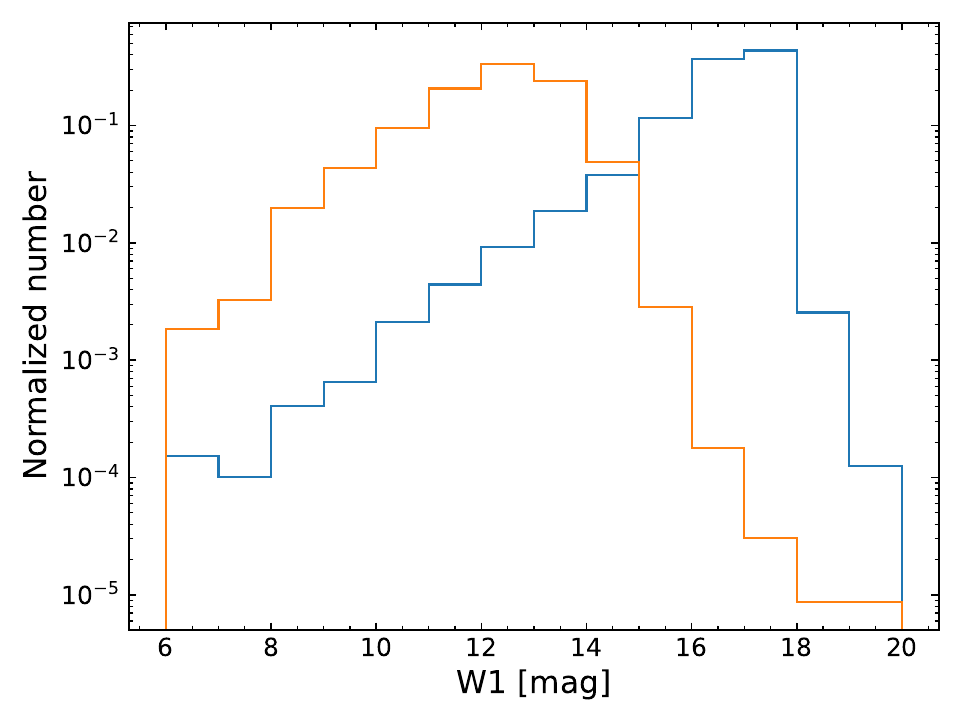}
    \caption{The blue line represents the magnitude distribution in COSMOS, and the orange line represents the magnitude distribution in our study region.}
    \label{fig:two_w1_df}
\end{figure}

\citet{2023AJ....165...36M} assessed the reliability and completeness of unTimely data in the COSMOS \citep{2007ApJS..172...86S} region at various magnitude intervals. This assessment was done by comparing the unTimely data with Spitzer data. Figure 3 in \citet{2023AJ....165...36M} illustrates that the 90\% reliability threshold for unTimely is 16.81 mag, while the 90\% completeness threshold is 17.14 mag in the W1 band. It is worth noting that the region under study is located in the Galactic plane, which has a relatively dense stellar field compared to the COSMOS region. Consequently, the depth of unTimely data varies between dense and non-dense stellar fields. Figure~\ref{fig:two_w1_df} presents the distribution of W1 magnitudes in a portion of the COSMOS region (coadd\_id=1497p015) and a portion of our study region (coadd\_id=2571m409) during epoch-0. It is evident from the figure that the detection depths differ between the two regions.

\begin{figure}
    \centering
        \includegraphics[width=\columnwidth]{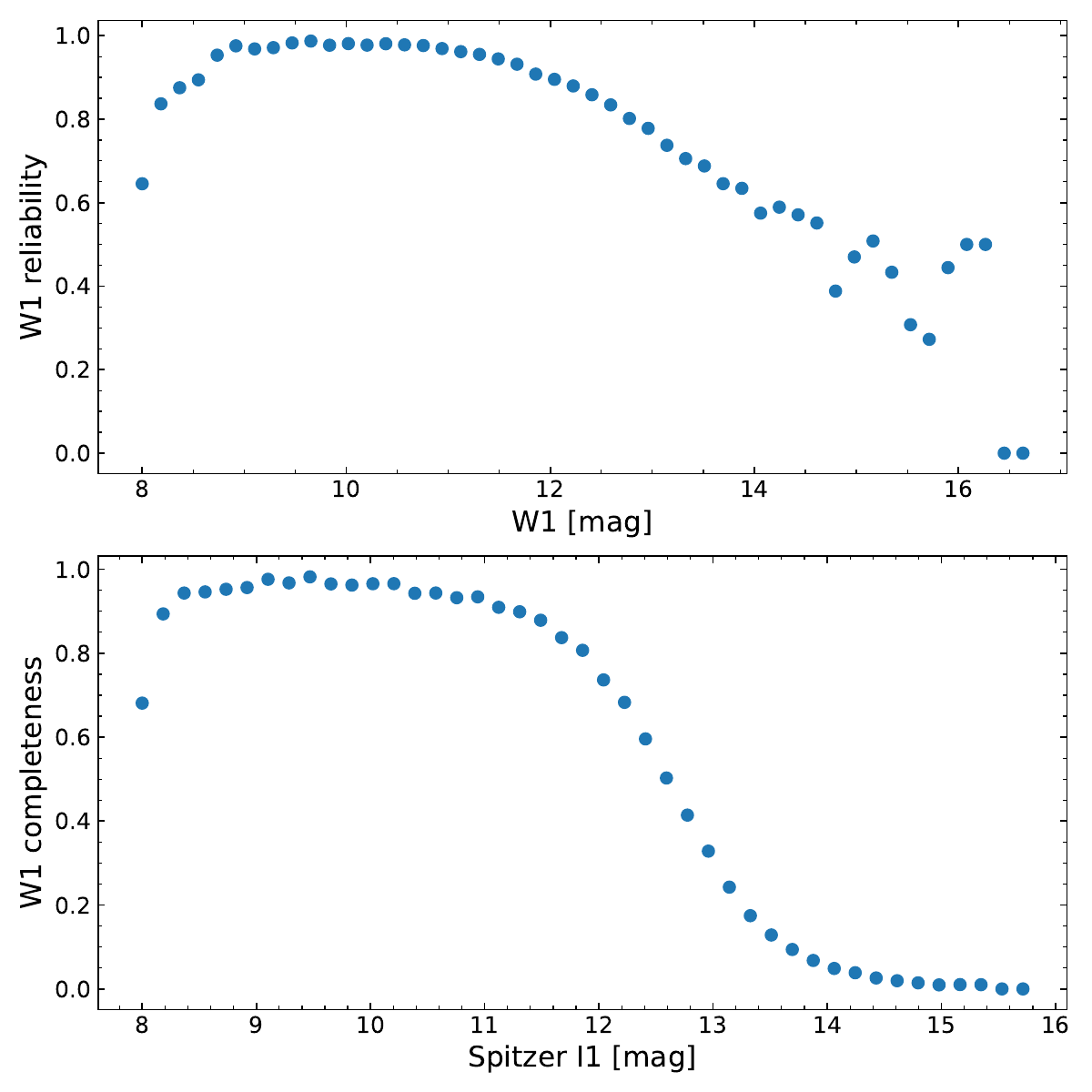}
    \caption{Completeness and reliability results for the unTimely Catalog in dense stellar field. The upper panel shows the reliability at different magnitudes, while the lower panel shows the completeness at different magnitudes.}
    \label{fig:acc_and_com_of_w1}
\end{figure}

In our study, we followed the methodology outlined in the work of \citet{2023AJ....165...36M} to compare a subset of the unTimely data within our study region with Spitzer data. The results of this comparison are presented in Figure~\ref{fig:acc_and_com_of_w1}. It is evident from the figure that, except for a decrease in reliability and completeness around 8 mag due to saturation, both reliability and completeness remain above 90\% below 11 mag. As a result, in Section \ref{Select method}, we selected 11 mag as the cutoff value in the second step of the filtering process.

\section{Further checking of the VVV data} \label{Further checking of the VVV data}
\begin{figure}
    \centering
        \includegraphics[width=\columnwidth]{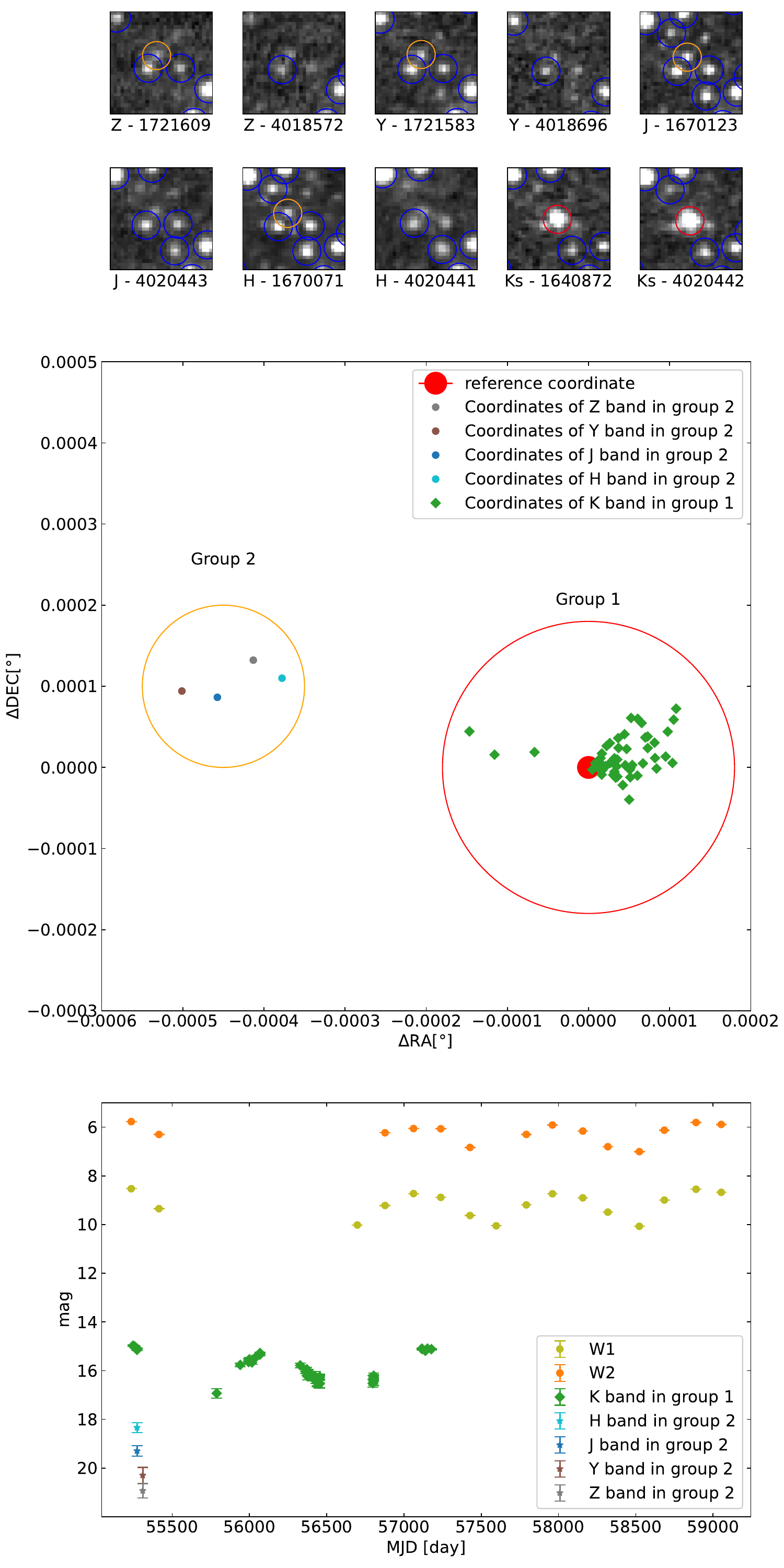}
    \caption{The upper panel shows the images of the ZYJHK$_\mathrm{s}$ band of the VVV of this source for different epochs. The text below each image writes the band and multiframe ID corresponding to the image. The red circle is the detection for group 1, and the orange circle is the detection for group 2, all within 1 arcsecond. The blue circles are unmatched data (due to the limitation of space, each band is only shown for 2 epochs, in fact, all images we will plotted during our visual inspection). The middle panel shows the distribution and grouping of the coordinates relative to unTimely for the individual bands and epochs of the individual bands for this source. The lower panel represents the unTimely light curve and the light curve of VVV in different groups. For this source we just keep the data in group 1.}
    \label{fig:example_of_vvv_crossmatch}
\end{figure}

Since most of our sources are infrared bright sources and the optical radiation of our sources may be weak, the bluer bands in the VVV data (such as Z and Y) may be matched to other sources. Not only that, because ZYJHK$_\mathrm{s}$ has more epochs in these bands, it is also possible to match with other sources in certain epochs in certain bands.
To solve this problem, we did the following checks. For a certain source, as shown in Figure~\ref{fig:example_of_vvv_crossmatch}, we plot the coordinates of all epochs for all bands of its VVV. These detections were self-classified by DBSCAN (Density-Based Spatial Clustering of Applications with Noise) algorithm. The radius of the parameter field and the minimum number of points are set to 0.0001$^\circ$ (about 1/3 arcsecond) and 1, respectively. And 1/3 arcsecond is about 1 pixel in the VVV images. After dividing these detections into groups, we did visually inspect the image information of each VVV band and epoch, the distribution of photometric coordinates around the reference coordinates, and the shape of the VVV light curve to determine which group (or groups) should be matched. In this example, we only selected the sources in group 1.

\section{Multi-epoch photometry and SED data} \label{multi-epoch photometry and sed data}
We present a part of unTimely multi-epoch photometry and VVV multi-epoch photometry for V1 source in Table \ref{tab:muti-epoch photometry data} and top 10 source's multi-band photometry in Table \ref{tab:muti-band data I} in this text. Full versions of these tables can be accessed online.

\begin{table}
	\centering
	\caption{Multi-epoch photometry data of MIR large-amplitude variable sources. "$\star$"  is used as the saturation flag. Here we only list a part of V1 source photomertry. The complete list is available in the online supplementary information.}
	\label{tab:muti-epoch photometry data}
	\begin{tabular}{lccccc} 
		\hline
		  Index & mjd & band & mag & magerr \\
		\hline
            \begin{filecontents*}{lightcurve.csv}
index,mjd,band,mag,err
V1,W1,55222.25,10.9511,0.0013
V1,W1,55400.13,10.6936,0.0011
V1,W1,55587.45,10.7785,0.0010
V1,W1,56684.97,11.4121,0.0018
V1,W1,56863.95,11.3207,0.0018
V1,W1,57049.35,10.6100,0.0011
V1,W1,57224.30,11.1256,0.0015
V1,W1,57414.78,10.7290,0.0011
V1,W1,57583.17,10.5542,0.0009
V1,W1,57779.87,10.7168,0.0012
V1,W1,57944.95,10.3907,0.0009
V1,W1,58144.22,10.6496,0.0013
V1,W1,58305.55,10.7742,0.0012
V1,W1,58509.63,11.1738,0.0017
V1,W1,58671.13,11.3299,0.0018
V1,W1,58875.30,11.0670,0.0017
V1,W2,55222.25,10.6107,0.0023
V1,W2,55400.13,10.2656,0.0017
V1,W2,55587.45,10.4317,0.0017
V1,W2,56685.03,11.2766,0.0039
V1,W2,56863.95,11.1824,0.0038
V1,W2,57049.28,10.4538,0.0022
V1,W2,57224.30,10.8964,0.0033
V1,W2,57414.78,10.4366,0.0021
V1,W2,57583.10,10.2634,0.0017
V1,W2,57779.87,10.3934,0.0021
V1,W2,57945.02,10.0993,0.0018
V1,W2,58144.22,10.3089,0.0024
V1,W2,58305.55,10.4273,0.0023
V1,W2,58509.63,10.8682,0.0030
V1,W2,58671.13,11.1487,0.0040
V1,W2,58875.30,10.8522,0.0034
V1,z,55281.11,12.7752,0.0008
V1,z,57140.21,12.8309,0.0013
V1,y,55281.10,12.3582,0.0008
V1,y,57140.21,12.4140,0.0012
V1,j,55270.16,12.1082,0.0009
V1,j,57114.02,*11.8829,0.0009
V1,h,55270.15,*12.0891,0.0009
V1,h,57114.01,*11.6821,0.0015
V1,ks,55227.29,11.8167,0.0012
V1,ks,55227.33,11.7132,0.0012
V1,ks,55269.12,*11.4644,0.0016
V1,ks,55270.15,11.5737,0.0011
V1,ks,55271.15,*11.4121,0.0016
V1,ks,55272.14,*11.4054,0.0016
V1,ks,55281.08,*11.2126,0.0016
V1,ks,55282.10,*11.1099,0.0015
V1,ks,55701.99,*11.4059,0.0018
V1,ks,55733.97,*11.0592,0.0015
...,...,...,...,...
\end{filecontents*}
\csvreader{lightcurve.csv}{}%
        {\csvcoli & \csvcoliii & \csvcolii & \csvcoliv & \csvcolv\\}\\
		\hline
	\end{tabular}
\end{table}

\begin{landscape}
\clearpage
\pagestyle{empty}
\setlength\LTleft{0pt}
\setlength\LTright{0pt}
\setlength\topmargin{-30pt}
\setlength\textwidth{702pt}
\setlength\textheight{38pc}
\begin{table*}
    \centering
    \scriptsize
    \caption{Multi-band of MIR large-amplitude variable sources I. zmag, ymag, jmag, hmag and K$_\mathrm{s}$mag are from VVV. I1mag, I2mag, I3mag and I4mag are from Glimpse. "$\star$"  is used as the saturation flag. Only the first 10 sources are shown. The complete list is available in the online supplementary information.}
    \label{tab:muti-band data I}
    \begin{tabular}{|p{0.4cm}|p{0.8cm}|p{0.9cm}|p{0.6cm}|p{0.6cm}|p{0.6cm}|p{0.6cm}|p{0.6cm}|p{0.6cm}|p{0.6cm}|p{0.8cm}|p{0.8cm}|p{0.8cm}|p{0.8cm}|p{0.8cm}|p{0.8cm}|p{0.8cm}|p{0.8cm}|p{0.8cm}|p{0.8cm}|p{0.8cm}|}
        \hline
        Index & RA & DEC & zmag & zmagerr & ymag & ymagerr & jmag & jmagerr & hmag & hmagerr & K$_\mathrm{s}$mag & K$_\mathrm{s}$magerr & I1mag & I1magerr & I2mag & I2magerr & I3mag & I3magerr & I4mag & I4mager\\
        \hline
        \begin{filecontents*}{sed.csv}
index,ra,dec,vvv_Zmag,vvv_Zmagerr,vvv_Ymag,vvv_Ymagerr,vvv_Jmag,vvv_Jmagerr,vvv_Hmag,vvv_Hmagerr,vvv_Kmag,vvv_Kmagerr,2mass_Jmag,2mass_Jmagerr,2mass_Hmag,2mass_Hmagerr,2mass_Kmag,2mass_Kmagerr,I1mag,I1magerr,I2mag,I2magerr,I3mag,I3magerr,I4mag,I4magerr,w1mag,w1magerr,w2mag,w2magerr,w3mag,w3magerr,w4mag,w4magerr
V1,175.94801,-62.83249,12.78,0.0008,12.36,0.0008,*12.11,0.0009,*12.09,0.0009,*11.46,0.0014,11.80,0.0250,11.37,0.0330,10.89,0.0260,10.93,0.0460,10.55,0.0530,10.22,0.0510,9.79,0.0420,10.76,0.0300,10.41,0.0250,--,--,--,--
V2,178.41579,-62.83414,15.98,0.0044,15.12,0.0037,14.20,0.0029,13.54,0.0026,13.08,0.0057,14.24,0.0510,13.44,0.0470,12.89,0.0450,13.33,0.0660,13.12,0.1050,12.71,0.1680,--,--,12.64,0.0380,12.50,0.0360,--,--,--,--
V3,178.21208,-62.84909,14.58,0.0018,13.75,0.0016,12.93,0.0014,*12.37,0.0012,11.98,0.0022,13.04,0.0230,12.39,0.0270,12.05,0.0270,11.92,0.0500,11.79,0.0620,11.57,0.0980,11.57,0.0620,11.64,0.0280,11.57,0.0310,10.87,0.1540,--,--
V4,175.67140,-62.82438,18.11,0.0231,16.72,0.0125,15.32,0.0064,14.21,0.0039,13.43,0.0062,15.11,0.0710,13.55,--,12.82,--,11.86,0.0480,11.35,0.0680,11.10,0.0840,10.68,0.0520,12.96,0.0330,12.84,0.0360,10.45,0.1270,--,--
V5,176.66073,-62.81652,--,--,--,--,15.07,0.0053,--,--,12.41,0.0032,14.73,0.0580,12.96,0.0570,11.96,0.0350,10.74,0.0440,10.05,0.0720,9.42,0.0470,8.59,0.0270,11.01,0.0240,10.30,0.0210,7.69,0.0330,5.00,0.0310
V6,177.77750,-62.72857,12.28,0.0006,11.89,0.0006,*11.59,0.0007,*12.00,0.0009,*11.21,0.0012,10.63,0.0280,10.34,0.0310,10.01,0.0260,9.86,0.0450,9.57,0.0470,9.29,0.0470,8.94,0.0300,10.30,0.0240,10.05,0.0210,9.08,0.0690,--,--
V7,177.04208,-62.44676,13.75,0.0013,13.05,0.0011,*12.30,0.0010,*12.21,0.0010,*11.39,0.0013,12.28,0.0250,11.59,0.0240,11.10,0.0230,11.03,0.0480,10.80,0.0440,10.59,0.0530,10.30,0.0310,10.99,0.0250,10.72,0.0240,9.45,0.0510,--,--
V8,177.33415,-61.42612,20.06,0.1331,18.88,0.0817,17.63,0.0471,15.34,0.0121,13.81,0.0116,--,--,--,--,--,--,12.47,0.0620,11.57,0.0510,10.80,0.0760,9.84,0.0300,13.10,0.0960,11.88,0.0360,8.70,0.0300,6.25,0.0510
V9,176.25583,-61.33202,19.48,0.0792,18.49,0.0586,17.05,0.0298,15.73,0.0178,14.78,0.0239,--,--,--,--,--,--,13.16,0.0600,12.16,0.0930,11.32,0.0750,10.18,0.0460,13.05,0.0770,11.92,0.0310,9.04,0.0530,--,--
V10,175.78930,-62.35374,--,--,--,--,19.06,0.1668,16.99,0.0443,15.66,0.0438,--,--,--,--,--,--,13.40,0.0700,12.42,0.1060,11.90,0.1240,11.16,0.0970,14.30,0.0550,12.91,0.0470,--,--,--,--
...,...,...,...,...,...,...,...,...,...,...,...,...,...,...,...,...,...,...,...,...,...,...,...,...,...,...,...,...,...,...,...,...,...,...
\end{filecontents*}
\csvreader{sed.csv}{}%
        {\csvcoli & \csvcolii & \csvcoliii & \csvcoliv & \csvcolv & \csvcolvi & \csvcolvii & \csvcolviii & \csvcolix & \csvcolx & \csvcolxi & \csvcolxii & \csvcolxiii & \csvcolxx & \csvcolxxi & \csvcolxxii & \csvcolxxiii & \csvcolxxiv & \csvcolxxv & \csvcolxxvi & \csvcolxxvii\\}\\
        \hline

    \end{tabular}
\end{table*}
\begin{table*}
    \centering
    \scriptsize
    \contcaption{Multi-band of MIR large-amplitude variable sources II. Jmag, Hmag and Kmag are from 2MASS. W1-W4 are from AllWISE. Only the first 10 sources are shown. The complete list is available in the online supplementary information.}
    \label{tab:muti-band data II}
    \begin{tabular}{|p{0.3cm}|p{1.1cm}|p{1.2cm}|p{1cm}|p{1cm}|p{1cm}|p{1cm}|p{1cm}|p{1cm}|p{1cm}|p{1cm}|p{1cm}|p{1cm}|p{1cm}|p{1cm}|p{1cm}|p{1cm}|}
        \hline
        Index & RA & DEC & Jmag & Jmagerr & Hmag & Hmagerr & Kmag & Kmagerr & W1mag & W1magerr & W2mag & W2magerr & W3mag & W3magerr & W4mag & W4magerr\\
        \hline
        \csvreader{sed.csv}{}%
        {\csvcoli & \csvcolii & \csvcoliii & \csvcolxiv & \csvcolxv & \csvcolxvi & \csvcolxvii & \csvcolxviii & \csvcolxix & \csvcolxxviii & \csvcolxxix & \csvcolxxx & \csvcolxxxi & \csvcolxxxii & \csvcolxxxiii & \csvcolxxxiv & \csvcolxxxv \\}\\
        \hline

    \end{tabular}
\end{table*}
\end{landscape}

\section{The light curves of FUor/EXor candidates} \label{The lightcurves of FUor/EXor candidates}
As mentioned before, we found 18 FUor candidates and one EXor candidate. Figure~\ref{fig:FE_light_curve1} shows their light curves.

\begin{figure*}
	\includegraphics[width=\textwidth]{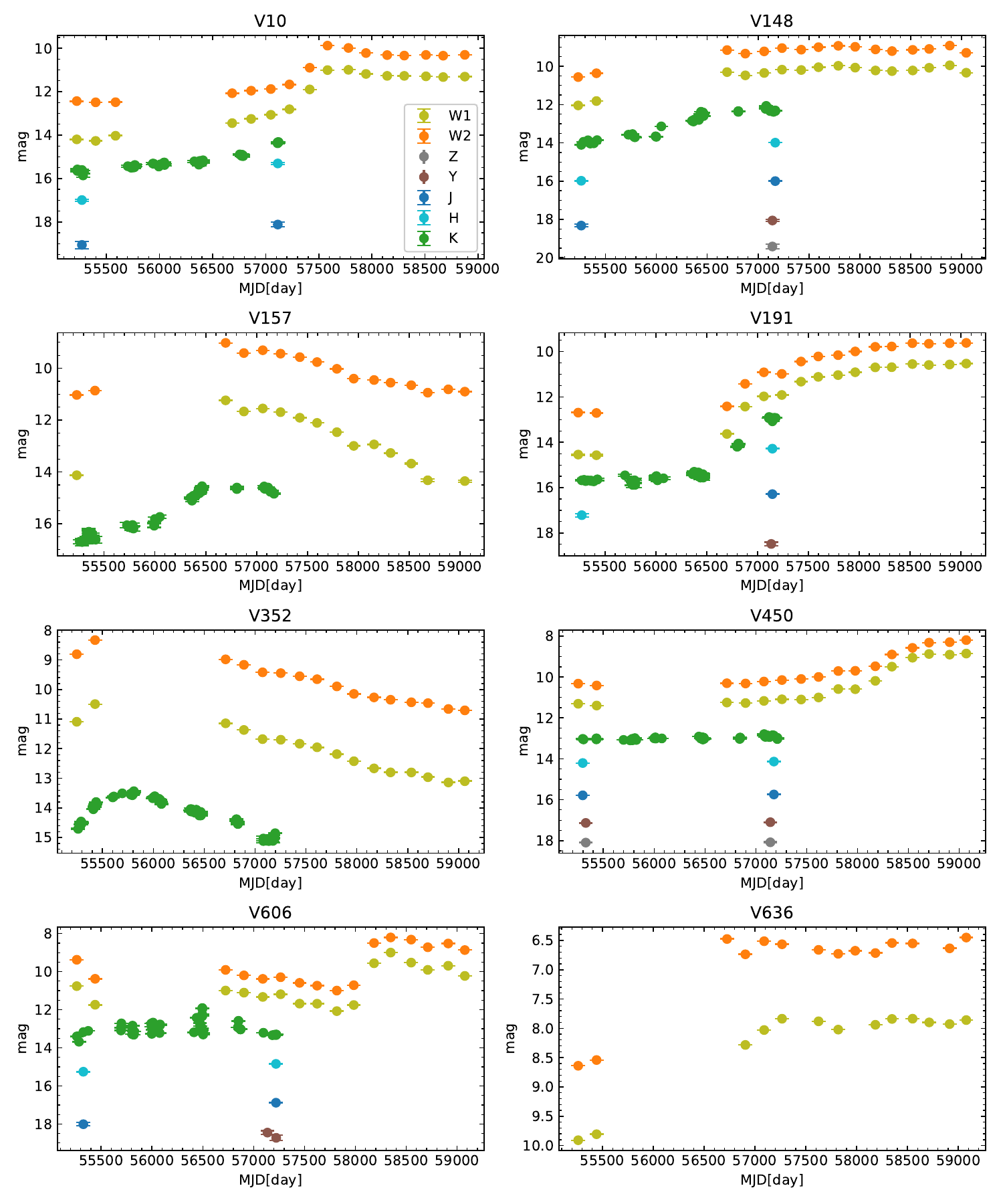}
    \caption{The light curve of the FUor candidates mentioned above. Colors denote different band, same as Figure~\ref{fig:lightcurve}
}
    \label{fig:FE_light_curve1}
\end{figure*}
\begin{figure*}
	\includegraphics[width=\textwidth]{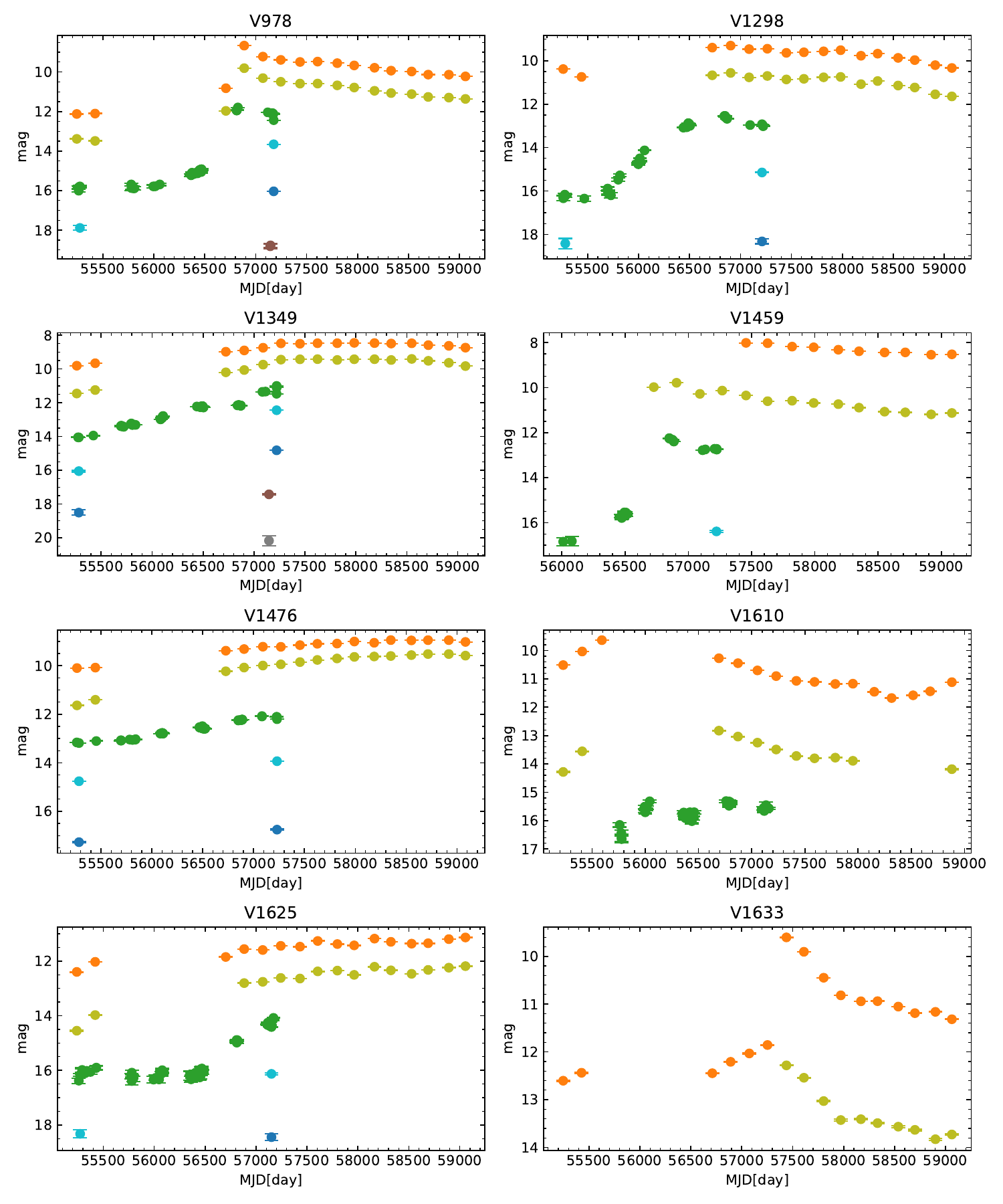}
    \contcaption{The light curve of the FUor candidates mentioned above. Colors denote different band, same as Figure~\ref{fig:lightcurve}
}
    \label{fig:FE_light_curve2}
\end{figure*}
\begin{figure*}
	\includegraphics[width=\textwidth]{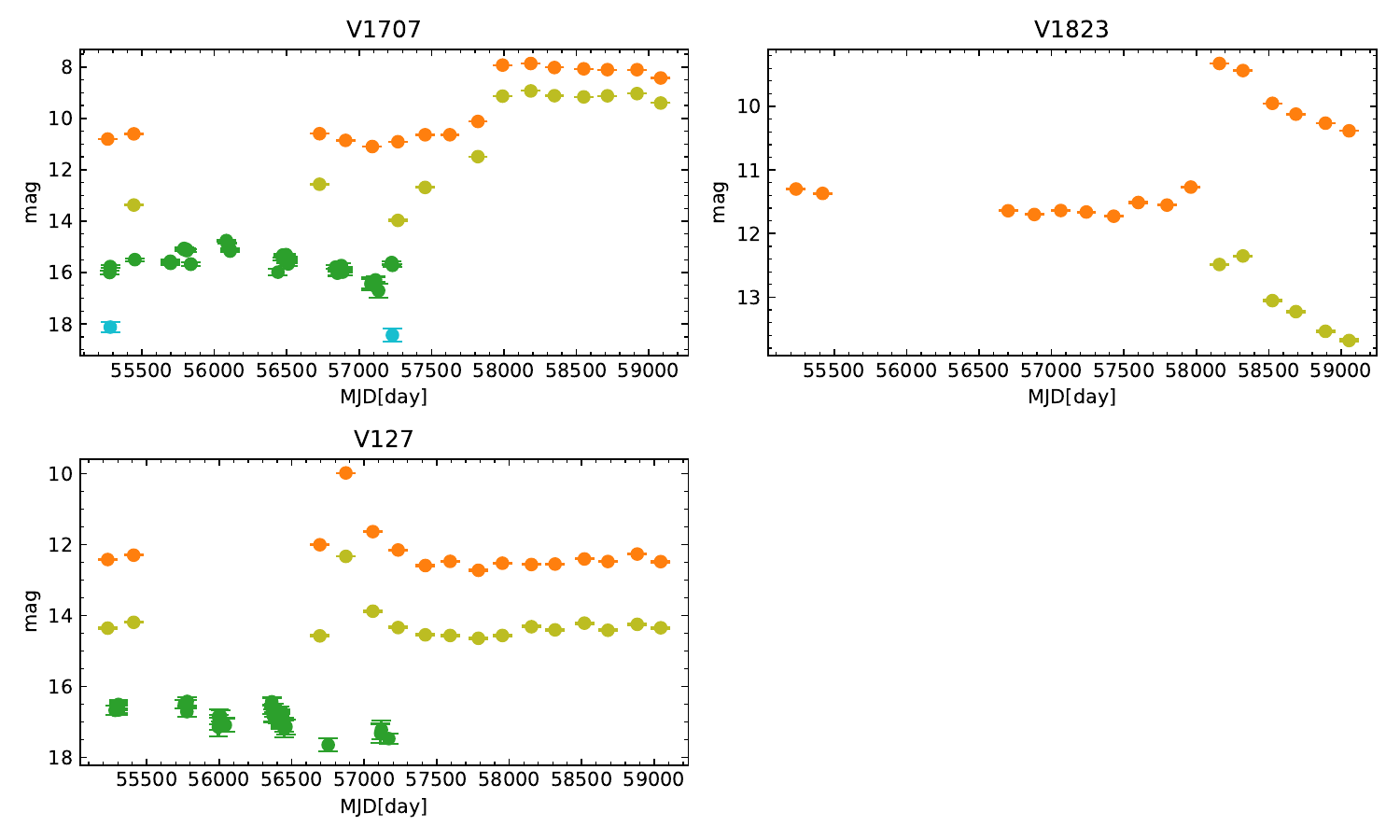}
    \contcaption{The light curves of FUor and EXor (V127) candidate. Colors denote different band, same as Figure~\ref{fig:lightcurve}
}
    \label{fig:FE_light_curve3}
\end{figure*}

\bsp	
\label{lastpage}
\end{document}